\documentclass[12pt,draftcls,onecolumn]{IEEEtran}
\usepackage{amsmath,amssymb,amsfonts,mathrsfs,bm}
\usepackage{amstext}
\usepackage{upgreek}
\usepackage{multicol}
\usepackage{graphicx}
\usepackage{paralist}
\usepackage{hyperref}
\usepackage[numbers,sort&compress]{natbib}
\usepackage{tabularx}
\usepackage{booktabs}
\usepackage{color}
\usepackage{multirow}
\usepackage{subfigure}
\usepackage{textcomp}

\usepackage{algorithm}
\usepackage{algorithmic}
\newtheorem{proposition}{Proposition}

\IEEEoverridecommandlockouts

\begin{document}
\title{Caching Policy for Cache-enabled D2D Communications by Learning User Preference}
\author{
{Binqiang Chen and Chenyang Yang }
%
\thanks{
	Binqiang Chen and Chenyang Yang are with the School of Electronics and Information Engineering, Beihang University, Beijing, China, Emails: \{chenbq,cyyang\}@buaa.edu.cn. Part of this work has been presented in IEEE VTC Spring 2017 \cite{chen17caching}.
	
}
}

\maketitle

\vspace{-10mm}\begin{abstract}
Prior works in designing caching policy do not distinguish \emph{content popularity} with \emph{user preference}.
In this paper, we illustrate the caching gain by exploiting individual user behavior in sending requests. After showing the connection between the two concepts, we provide a model for synthesizing user preference from content popularity. We then optimize the caching policy with the knowledge of user preference and active level to maximize the offloading probability  for cache-enabled device-to-device communications, and develop a low-complexity algorithm to find the solution. In order to learn user preference, we model the user request behavior resorting to probabilistic latent semantic analysis, and learn the model parameters by expectation maximization algorithm. By analyzing a Movielens dataset, we find that the user preferences are less similar, and the active level and topic preference of each user change slowly over time. Based on this observation, we introduce a prior knowledge based learning algorithm for user preference, which can shorten the learning time. Simulation results show remarkable performance gain of the caching policy with user preference over existing policy with content popularity, both with realistic dataset and synthetic data validated by the real dataset.
\end{abstract}

\begin{IEEEkeywords}
User preference, Content popularity, Caching policy, D2D, Machine learning, data analysis.
\end{IEEEkeywords}

\section{Introduction}

Caching at the wireless edge has become a trend for content delivery \cite{Procach14,LHui14,dongcaching16}, which can improve network throughput and energy efficiency as well as user experience dramatically \cite{Golrezaei.TWC,Dong}.

Owing to the small storage size of each node at the wireless edge, say  base station (BS) or user device, caching in a proactive manner is critical to achieve the performance gain, where future user demand statistics is exploited \cite{Procach14,leconte2016placing}.
In wireless networks, the contents can be precached at each BS  \cite{dongcaching16,Procach14} or even directly pushed to user device \cite{LHui14}. By caching at BS, backhaul traffic can be offloaded and backhaul cost can be reduced. By caching at user, wireless traffic can be further offloaded from peak time to off-peak time \cite{Ali13}. To boost the cache hit rate by precaching contents at each user that has very limited cache size, cache-enabled device-to-device (D2D) communications and coded-multicast strategy are proposed \cite{Golrezaei.TWC,Procach14,Ali13}.


To heap the proactive caching gain, various caching policies have been optimized with diverse objectives for different networks.
Most existing works assume known content popularity, defined as the probability distribution that every content in a catalog is requested by all users. For example, the policies for caching at BSs were optimized to minimize average download delay in \cite{niki2012femto} and to maximize coverage probability in \cite{B2015optimal}. The policies for caching at users were optimized to maximize the offloading gain of cache-enabled D2D networks in \cite{JMY.JSAC,chen2017energy}. Coded caching policy was optimized to maximize the average fractional offloaded traffic
and average ergodic rate of small-cell networks in \cite{cui17analysis}. In these works, every user is assumed to request files according to {content popularity}. However, in practice a user actually sends requests according to its own preference, which may not be identical to content popularity.
Noticing such fact, caching policies at the user groups with different group popularity were optimized  in \cite{guo2015cooperative} to minimize the average delay of cache-enabled D2D communications.

To implement above-mentioned proactive caching policies, content popularity needs to be predicted \cite{Gha2016provab}. Popularity prediction has been investigated for diverse applications such as advertisement, where content popularity is defined as the accumulated number of requests every content in a catalog received or the request arrival rate for every content. Numerous methods have been proposed  \cite{tatar2014survey}.
 By using these prediction methods, the content popularity defined with probability in  \cite{niki2012femto,Procach14,Golrezaei.TWC,LHui14,Dong,dongcaching16,B2015optimal,cui17analysis,JMY.JSAC,guo2015cooperative,chen2017energy} can be obtained as a ratio of the number of requests for each file to the number of all requests.
In cellular networks, the number of users in a cell is much less than that in a region covered by a content server, and a mobile user may send requests to more than one BS. Since popularity depends on the group of users who send requests, the local popularity in a cell may differ from the global popularity observed at a server. Designing proactive caching policy at wireless edge with predicted global popularity leads to low cache hit rate  \cite{leconte2016placing}.

Optimizing caching policy for a BS should be based on the local popularity, which is formed by the users in a cell. In \cite{Blasco14learning}, local popularity was predicted as the number of requests received for each file at a small BS divided by the observation time, i.e., the request arrival rate. Then, the popularity is adjusted with a perturbation term, which is learned by applying multi-arm bandit algorithm, and finally the predicted popularity with perturbation term is used for caching policy optimization. The prediction is based on the cumulative growth method  \cite{tatar2014survey} and under the assumption that only the requests for already cached files can be observed, hence the learning algorithm is slow.
In \cite{bacstuug2015big}, the content popularity was predicted with a real dataset measured from cellular network. By converting the number of requests received at each BS into rating, the local popularity was predicted by a widely-used collaborative filtering technique, matrix factorization \cite{ekstrand2011collaborative}, with which the files with largest ratings are cached at the BS.

\vspace{-2mm}\subsection{Motivation and Contributions}

Since caching at wireless edge is motivated by the observation that the majority of requests are initiated for a minority of popular contents, a large body of priori works assume that all users send their requests according to content popularity.  The following facts, which are widely recognized in the communities studying recommendation problem and analyzing user behavior with real data, are largely overlooked in the community of studying caching at wireless edge: (i) as a demand statistic of multiple users, content popularity can not reflect the personal preference of each individual user \cite{Recommendation2016}, (ii) only a small portion of users are active in creating traffic \cite{paul2011understanding}. In fact, existing works do not differentiate content popularity from user preference. In practice, user preferences are heterogeneous although they may exhibit similarity to a certain extent. The caching policy designed under unrealistic assumptions inevitably yields performance loss.

In this paper, we investigate the gain of optimizing caching policy by learning user preference and active level over content popularity. To this end, we take cache-enabled D2D communications as an example system and offloading probability as an example objective. Because there are different definitions in the domains of computer science and wireless communications,
we first define user preference and active level as well as content popularity to be used in this paper, and provide a probabilistic model to synthesize user preference from content popularity by introducing similarity among user preferences. We then formulate an optimization problem with known user preference and active level to maximize offloading probability. Since the problem is NP-hard, a local optimal algorithm is proposed to reduce the complexity, which achieves at least 1/2 optimality. In order to learn user preference and active level, we model user request behavior resorting to probabilistic latent
semantic analysis (pLSA) originally proposed
for natural language processing \cite{hofmann1999prob}, whose model parameters can be learned using
approximate inference methods such as expectation maximization (EM) \cite{demp1977maximum}. With the help of pLSA model to decompose the user behavior into different components and based on the observation from analyzing a real dataset that active level and topic preference change slowly over time, we provide a prior knowledge based algorithm, which  can quickly learn user preference.

The major contributions of this paper are summarized as follows:
\begin{itemize}
 	\item We illustrate the caching gain of exploiting user preference and active level by optimizing an caching policy for D2D communications, and predict the behavior of each individual user by estimating model parameters of pLSA. We introduce a prior knowledge based algorithm to learn user preference, which shows the potential of transfer learning.
 	\item We characterize the connection between content popularity and user preference, provide a probabilistic model for synthesizing user preference
from content popularity, and validate the method by the MovieLens 1M dataset \cite{harper2016movielens}. We analyze the relation between file catalog size and the number of users, the statistics of active level and topic preference of each user, and the user preference by the real dataset, which are critical for the caching gain.
 	\item Simulation results with both synthesized data and MovieLens dataset show remarkable performance gain of the caching policy with user preference over that with local popularity, no matter the user demands are assumed known or learned from historical requests.
\end{itemize}

\vspace{-2mm}
\subsection{Related Works}
Considering that the tastes of different users are not identical, caching policies were optimized to minimize the average delay of cache-enabled D2D communications in \cite{zhang2016clustered} and to maximize the cache hit rate of mobile social networks in \cite{wu2016semigradient}, by assuming user preferences as Zipf distributions with different ranks. However, both works assume that all users have the same active level, do not validate the assumption for user preference, and do not show the gain over caching with popularity. Until now, there exists no method to synthesize user preference validated by real dataset, and the gain of caching with user preference is unknown.

There are few works that consider the relation between content popularity and user preference.
In \cite{bastug2013proactive}, local popularity is computed as a weighted average of preferences for the users associated with each BS, where the weight is the number of requests sent by each user and the user preference was assumed as uniform distribution. Then, the most popular files at each BS were cached. Differing from this early work, we characterize the connection between the collective and the individual user request behavior in a probabilistic framework, and illustrate the gain of exploiting user preference over popularity by optimizing a caching policy.

These priori works assume known user preference \cite{zhang2016clustered,wu2016semigradient,bastug2013proactive}.
To facilitate proactive caching, user preference needs to be predicted, which is a key task in recommendation problem. Collaborative filtering is the most commonly used technique to predict user preference\cite{ekstrand2011collaborative}, which can be mainly classified into memory based method including user-based and item-based approaches, and model based method that is based on models with parameters estimated from historical records  \cite{shi2014collaborative}. Typical model based methods employ matrix factorization, latent
Dirichlet allocation \cite{LDA}, and pLSA \cite{hofmann2004latent} as models \cite{ekstrand2011collaborative}. For recommendation problem, user preference is defined as the rating that a user gives for a file, such as $0 \sim 5$ or simply ``like'' and ``dislike''. Most collaborative filtering methods predict the ratings for unrated contents of each user, which however cannot be used in optimization for wireless caching. To optimize caching policy in wireless edge, where various metrics are in statistical sense \cite{niki2012femto,Procach14,LHui14,Golrezaei.TWC,Dong,chen2017energy,dongcaching16,B2015optimal,cui17analysis,JMY.JSAC,guo2015cooperative}, user preference needs to be defined in probabilistic form, but there is no widely-accepted approach to map the rating into probability. In this paper, we resort to pLSA to model and predict user preference, which is originally developed for classification in automatic indexing \cite{hofmann1999prob} and then is applied to predict ratings in \cite{hofmann2004latent}.

The rest of the paper is organized as follows. Section \ref{poppre} provides the relation between content popularity and user preference, and a model to synthesize user preference. Section \ref{sec:caching} optimizes the caching policy with known user preference. Section \ref{sec:predict} presents the learning algorithms. Section \ref{dataset} analyzes the statistics of user demands from and validate the synthetic model by a MovieLens dataset. Section \ref{sec:simulation} provides simulation results. Section \ref{sec:conclusion} concludes the paper.
	
\section{Content Popularity and User Preference}
\label{poppre}
In this section, we first define content popularity, user preference and active level to be used in the following, show their connection, and then provide a probabilistic model with a free parameter to synthesize user preference.
\vspace{-3mm}
\subsection{Definition and Relationship}
Consider a content library $\mathcal{F} = \{{\rm f}_1,{\rm f}_2,...,{\rm f}_{F}\}$ consisting of $F$ files that $K$ users in an area  may request, where ${\rm f}_f$ denotes the $f$th file.

{\em Content popularity} is defined as the probability distribution that each file in the library is requested by all users, denoted as  ${\bf p} = [p_1,p_2,...,p_{F }]$, where $p_{f} \triangleq P({\rm f}_f)$ is the probability that ${\rm f}_f$ is requested, $\sum_{f=1}^{F} p_f = 1$, $p_{f}\in [0,1]$, and $1 \leq f \leq F$. If the area only consists of a single cellular cell, then ${\bf p}$ is called local popularity.

{\em User preference} is defined as the conditional probability distribution that a user requests a file given that the user sends a request, denoted as ${\bf q}_k= [q_{1|k},q_{2|k},...,q_{F|k }]$ for the $k$th user (denoted as ${\rm u}_k)$), where $q_{f|k} \triangleq P({\rm f}_f|{\rm u}_k)$ is the conditional probability that the $k$th user requests the $f$th file when the user sends a file request, $\sum_{f=1}^{F} q_{f|k}= 1$, $q_{f|k}\in [0,1]$, $1 \leq f \leq F$ and $1 \leq k \leq K$. We use matrix $\textbf{Q}= (q_{f|k})^{K \times F}$ to denote the preferences of all users, where $(q_{f|k})^{K \times F}$ represents a matrix with $K$ rows and $F$ columns and $q_{f|k}$ as the element in the $k$th row and $f$th column.

{\em Active level} is defined as the probability that a request is sent by a user, denoted as $w_k \triangleq P({\rm u}_k)$ for the $k$th user, which reflects how active the user is, where $\sum_{k=1}^{K} w_k = 1$ and $w_k \in [0,1]$. Then, the vector ${\bf w} = [w_1,w_2,...,w_K]$ denotes the active level distribution of the $K$ users.

Content popularity ${\bf p}$ reflects the
collective request behavior of a group of users, while ${\bf q}_k$ and ${w_k}$ characterize the individual request behavior of the $k$th user.
{To show their connection, we consider a $K \times F$ user-content matrix \cite{ekstrand2011collaborative}, where each of its element $n_{k,f}$ represents the number of requests sent by ${\rm u}_k$ for ${\rm f}_f$. Denote $N = \sum_{k=1}^{K}\sum_{f=1}^{F} n_{k,f}$ as the overall number of requests sent by all the $K$ users for all the $F$ files, $n_f = \sum_{k=1}^{K} n_{k,f}$ as the total number of requests sent by all users for ${\rm f}_f$ (i.e., the sum of all elements in the $f$th column), and $n_k = \sum_{f=1}^{F} n_{k,f}$ as the total number of requests sent by ${\rm u}_k$ for all files (i.e., the sum of all elements in the $k$th row). Considering that $n_f/N$, $n_k/N$ and $n_{k,f}/n_k$ are respectively multinomial distributions with $F$, $K$, and $F$ parameters, it is not hard to show that they are respectively the maximum likelihood estimate of $p_f$, $w_k$ and $q_{f|k}$. From their definitions, we have
\begin{equation}
\label{equ.connection}
\sum_{k=1}^{K}  \underbrace{{n_k}/{N}}_{w_k}  \underbrace{{n_{k,f}}/{n_k} }_{q_{f|k}}  = \sum_{k=1}^{K} \frac{n_{k,f}}{N} = \underbrace{{n_f}/{N}}_{p_f},
\end{equation}}
and hence each element of ${\bf p}$ can be expressed as the average of user preferences weighted by their active levels,
\begin{equation}
	\label{p_f} \textstyle
	p_f = \sum_{k=1}^{K} w_k q_{f|k}, \quad 1 \leq f \leq F.
\end{equation}

 In practice, users have different tastes, i.e., ${\bf q}_k \neq {\bf q}_m$, and hence $q_{f|k} \neq p_f$, despite that users may have similar preferences, say for popular contents. Besides, not all users send requests with equal probability.
We can use cosine similarity to reflect the similarity of preferences between two users, which is frequently used in collaborative
filtering  \cite{ekstrand2011collaborative} and defined as
\begin{equation}
\label{cos_sim}\textstyle
\text{sim}({\bf q}_k,{\bf q}_m) = \frac{ \sum_{f=1}^{F} q_{f|m} q_{f|k}}{\sqrt{\sum_{f=1}^{F} q_{f|m}^2 \sum_{f=1}^{F} q_{f|k}^2}}.
\end{equation}
To show the similarity among $K$ users, we can define average cosine similarity as
\begin{equation}
\label{ave_sim}\textstyle
\mathbb{E}_{k,m} [ \text{sim}({\bf q}_k,{\bf q}_m) ]= \frac{2}{K(K-1)}\sum_{k,m}\frac{ \sum_{f=1}^{F} q_{f|k} q_{f|m}}{ \sqrt{\sum_{f=1}^{F} q_{f|k}^2 \sum_{f=1}^{F} q_{f|m}^2}}.
\end{equation}

\vspace{-4mm}
\subsection{Modeling and Synthesizing User Preference}

{Content popularity} can be modeled as a Zipf distribution according to the analyses for many real datasets \cite{gill2007youtube,bacstuug2015big,cha2007tube}. The probability that the $f$th file is requested by all users is
\begin{equation}
\label{content_p} \textstyle
p_f = {f^{-\beta}}/{\sum_{j=1}^{F} j^{-\beta} }, \quad1 \leq f \leq F,
\end{equation}
where the files in the library are indexed in descending order of popularity, and the content popularity is more skewed with larger value of $\beta$.

{User preference} model obtained from real datasets is unavailable in the literature so far. Inspired by the method in \cite{leconte2016placing} to synthesize local popularity of a cell from that of a core network, we represent users and files in a shared one-dimensional latent space, which bears the same spirit as latent factor model widely applied in collaborative
filtering \cite{LDA,CTM}. To connect with  content popularity, we model user preference from the following generative process:
\begin{itemize}
	\item ${\rm u}_k$ is associated with a feature value $X_k$, which is randomly selected from $[0,1]$.
	\item ${\rm f}_f$ is associated with a feature value $Y_f$, which is again chosen uniformly from $[0,1]$.
	\item The joint probability that the $f$th file is requested by the $k$th user is given by
\begin{equation}
\label{user_p}\textstyle
P({\rm u}_k, {\rm f}_f) = w_kq_{f|k} = p_f \frac{g(X_k,Y_f)}{\sum_{k'=1}^{K} g(X_{k'},Y_f)},
\end{equation}
and then ${\rm u}_k$'s active level is ${w}_k = \sum_{{\rm f}_f \in \mathcal{F}}P({\rm u}_k , {\rm f}_f )$ and its preference is ${q}_{f|k} = P({\rm u}_k , {\rm f}_f )/w_k$, where $g(X_k,Y_f) \in [0,1]$ is a kernel function used to control the correlation between the $k$th user and the $f$th file.
When $g(X_k,Y_f) = 0$, the $f$th file will never be requested by the $k$th user. When $g(X_k,Y_f) = 1$, the file is a preferred file of the user.
\end{itemize}

Intuitively, the value of $X_k$ can be interpreted as the likelihood that the $k$th user prefers a topic, and the value of $Y_f$ can be interpreted as the likelihood that the $f$th file belongs to a topic.
 Various kernel function can be applied, e.g., Gaussian, logarithmic and power kernels. To control the average similarity among the user preferences by introducing a parameter $\alpha$ in kernel function, we choose power kernel with expression $g(X_k,Y_f) = (1-|X_k-Y_f|)^{(\frac{1}{\alpha^3}-1)} \in [0,1]$ ($0<\alpha \leq 1$), which exhibits a linear relation between $\mathbb{E}_{k,m} [ \text{sim}({\bf q}_k,{\bf q}_m) ]$ and $\alpha$ in a wide range.

{\bf Remark 1:}
When $\alpha = 1$, $g(X_k,Y_f) = 1$ for $\forall k,f$. Then, we can see from \eqref{user_p} that all user preferences are identical and equal to the content popularity, and then from  \eqref{ave_sim} we can obtain $\mathbb{E}_{k,m}[\text{sim} ({\bf q}_k,{\bf q}_m) ]=1$. When $\alpha \rightarrow 0$, $g(X_k,Y_f) \rightarrow 0$ for $X_k \neq Y_f$, and $g(X_k,Y_f) = 1$ only for $X_k = Y_f$. Because $X_k$ and $Y_f$ are uniformly chosen from $[0,1]$, i.e., $\mathbb{P}(X_k = Y_f) \rightarrow 0$, we have $g(X_k,Y_f)g(X_{k'},Y_f) \rightarrow 0, k \neq k'$. Then, according to \eqref{user_p} and \eqref{ave_sim}, $\mathbb{E}_{k,m}[ \text{sim} ({\bf q}_k,{\bf q}_m) ] \rightarrow 0$, i.e., no user has the same preference.
When $\alpha$ is small, $g(X_k,Y_f)$ is low, which means that the number of interested files of ${\rm u}_k$ is small. Because each user randomly chooses feature
values, the interested file sets among different users
are less overlapped, and hence the average similarity among the users is low.

This probabilistic model is appropriate for generating synthetic data of user preference and active level, which differs from pLSA that can be used for predicting the individual behavior. This model will be validated later by a real dataset and the parameter will be fitted.

\section{Caching Policy Optimization: an Illustration}
\label{sec:caching}
In this section, we illustrate how to optimize the caching policy with known request behavior of each individual user. For comparison, we also provide the corresponding caching policy optimization problem with known content popularity, whose solution reflects the existing policy in literature. To focus on the performance gain brought by distinguishing user preference from content popularity, we consider a simple objective: the offloading gain of D2D communications. To reduce the time complexity in finding the solution, we provide a local optimal algorithm.
	
\vspace{-2mm}\subsection{System Model}
\label{sec:system model}
 Multiple BSs in the area are connected to core network via backhaul to serve the $K$ uniformly distributed users, which constitutes a set of users $\mathcal{U} = \{{\rm u}_1,{\rm u}_2,...,{\rm u}_K\}$ that request the files in content library $\mathcal{F}$. Assume that each file is with same file size, but the results are applicable for general case with different sizes \cite{leconte2016placing} by dividing each file into chunks of approximately equal size.\footnote{We can also formulate another optimization problem with different file sizes, which can be shown as a knapsack problem.} Each single-antenna user has a local cache to store $M$ files, and can act as a helper to share files via D2D link.
To provide high rate  transmission with low energy cost at each user device, we consider a user-centric D2D communication protocol as in \cite{chen2017energy}. A helper can serve as a D2D transmitter and send its cached files to a user only if their distance is smaller than a collaboration distance  $r_{\rm c}$, which reflects the coverage of the helper. Each BS is aware of the cached files at and the locations of the users, and coordinates the D2D communications.

Proactive caching consists of content delivery phase and content placement phase.

In content delivery phase, each user requests files according to its own preference. If a user can find its requested file in local cache (i.e., fetching locally), it directly retrieves the file with zero delay. If not, the user sends the request to a BS. If the BS finds
that the file is cached in the local caches of helpers adjacent to the user, it informs the request to the closest helper, and then a D2D link is established between
the user and the closest helper (i.e., fetching via D2D link). Otherwise, the BS fetches the file via backhaul to serve the user. For simplicity, both fetching locally and via D2D link are called fetching via D2D links in the sequel.

Denote the file requests matrix after a period as ${\bf N}=(n_{k,f})^{K \times F}$ (i.e., the
user-content matrix), where $n_{k,f} \ge 0$ is the number of requests from ${\rm u}_k \in \mathcal{U}$ to ${\rm f}_f \in \mathcal{F}$. 
Assume that a central processor (CP) can record the requests history of users. To predict user preference, the CP needs to be deployed in the mobile core, such that most requests of users can be recorded. To determine where to deploy CP, one needs to consider both coverage area and computational cost.

In content placement phase, the CP learns the user preferences $\bf {Q}$ and active levels ${\bf {w}}$ from the requests history $\bf N$, and then optimizes the caching policy for users and informs the cached files of the users to the BSs.  We consider deterministic caching policy,\footnote{We do not consider probabilistic caching policy, which is designed under the assumption that a group of nodes share the same caching distribution \cite{niki2012femto,B2015optimal,cui17analysis,JMY.JSAC,chen2017energy,guo2015cooperative}, and hence is not appropriate for a system with heterogeneous user preferences. } denoted as a vector ${\bf c}_k = [c_{k,1},c_{k,2},...,c_{k,F}]$ for the $k$th user, where $c_{k,f}=1$ if  ${\rm f}_f$ is cached at ${\rm u}_k$, $c_{k,f}=0$ otherwise, and $\sum_{f=1}^{F} c_{k,f}\leq M$.  Denote the caching policy for all users as $\textbf{C}= (c_{k,f})^{K \times F}$. After being informed about the files to be cached at the users in its cell, a BS fetches the files from the core network and refreshes the caches of the users during the off-peak time, say every several hours, noticing that traffic load varies on the timescale of hours as measured by real cellular data \cite{paul2011understanding}.
Due to user preference similarity, some users may have common
contents to precache  as shown later, which can be pre-downloaded by the BS via multicast to reduce wireless traffic.


\vspace{-2mm}\subsection{Caching Policy Optimization with Individual Request Behavior}
%
We use {\em offloading probability} to reflect the offloading gain introduced by cache-enabled D2D communications, defined as the probability that a user can fetch the requested file via D2D links, which represent the average ratio of requests able to be offloaded.

When optimizing caching policy in the content placement phase, it is hard to know where a mobile user will be located in the content delivery phase. Therefore, it is hard to know when and how long the users will contact. Fortunately, data analysis shows that users always periodically reappear at the same location with high probability \cite{hsu2007model}. Consequently, it is reasonable to assume that the contact probability is known {\em a priori} \cite{wu2016semigradient}. Let $\textbf{A}= (a_{i,j})^{K \times K}$ represent the contact probability among users, where $a_{i,j} \in [0,1]$ is the probability that the distance between the $i$th user and the $j$th user is less than $r_c$. When all users do not move, $a_{i,j}=0$ or 1.

In D2D communications, adjacent helpers may have overlapped coverage. Since different helpers need to serve different groups of users, which depend not only on $r_c$ but also on the cached files at the adjacent helpers, the ``local popularity'' observed at a helper differs from that observed at another helper and relies on the caching policy. As a result, the caching policy can not be designed based on the ``local popularity''.

Denote $p^{\rm d}_{k,f}({\bf A,C})$ as the probability that the $k$th user can fetch the $f$th file via D2D links given contact probability ${\bf A}$ and caching policy {\bf C}. The complementary probability of $p^{\rm d}_{k,f}({\bf A,C})$ is the probability that the $f$th file is not cached at any users in proximity to the $k$th user, which can be derived as $\prod_{m=1}^{K} (1- a_{k,m}c_{m,f})$. Then, we can obtain
the offloading probability as
\begin{equation}
\label{p_off}
p_{\rm off} ({\bf Q,w,A,C}) = \sum_{k=1}^{K} w_k  \sum_{f=1}^{F} q_{f|k} p^{\rm d}_{k,f}({\bf A,C}) =\sum_{k=1}^{K} w_k \sum_{f=1}^{F} q_{f|k} \left(1-\prod_{m=1}^{K} (1- a_{k,m}c_{m,f}) \right).
\end{equation}



With known user preference and active level, the caching policy can be optimized to maximize the offloading probability by solving the following problem,
\begin{subequations}
	\begin{align}\textstyle
	\textbf{P1}: \quad& \textstyle    \max_{c_{m,f}}\,\, &&  \textstyle  p_{\rm off} ({\bf {Q},{w},A,C}) \label{equ.opt1} \\
	&   \textstyle s.t.  &&  \textstyle \sum_{f=1}^{F} c_{m,f} \leq M, c_{m,f} \in \{0,1\},   1 \leq m \leq K, 1 \leq f\leq F.  \label{equ.opt1_const}
	\end{align}
\end{subequations}

{\bf Remark 2:} If all users are with equal active level and equal preference, then \eqref{p_off}  becomes
\begin{equation}\textstyle\label{poff_base}
p_{\rm off} ({\bf Q,w,A,C})= \frac{1}{K} \sum_{f=1}^{F} p_{f} \sum_{k=1}^{K} p^{\rm d}_{k,f}({\bf A,C}) \triangleq p^{\rm pop}_{\rm off} ({\bf p,A,C}).
\end{equation}


{\bf Remark 3:} If the collaboration distance $r_c \rightarrow \infty$, then $a_{k,m}=1$, and \eqref{p_off} becomes
\begin{equation}
\begin{aligned}\textstyle
p_{\rm off} ({\bf Q,w,A,C}) & = \sum_{f=1}^{F} \left(1-\prod_{m=1}^{K} (1- c_{m,f}) \right) \sum_{k=1}^{K} w_k q_{f|k} = \sum_{f=1}^{F} \left(1-\prod_{m=1}^{K} (1- c_{m,f}) \right) p_f.\nonumber
\end{aligned}
\end{equation}
It is easy to show that $p_{\rm off} ({\bf Q,w,A,C})=p^{\rm pop}_{\rm off} ({\bf p,A,C})$ in this extreme case where D2D links can be established between any two users in the area even with heterogeneous user preferences.

If the assumptions in the two remarks hold, then the ``local popularity'' observed at every helper will be identical, which is equal to the content popularity of the area with the $K$ users.
In practice, the assumptions are not true, hence the caching gain from exploiting user preference differs from exploiting content popularity.

With known content popularity, the caching policy is optimized by maximizing $p^{\rm pop}_{\rm off} ({\bf p,A,C})$ in \eqref{poff_base} under constraint \eqref{equ.opt1_const}, called problem \textbf{P2},\footnote{The solution of \textbf{P2} slightly differs from \cite{JMY.JSAC,chen2017energy}, where  the future user location is exactly known in \cite{JMY.JSAC} and completely unknown in  \cite{chen2017energy} when optimizing caching policy, where the contact probability is known  in \textbf{P2}.} which is actually a special case of \textbf{P1}.

By setting the contact probability $a_{i,j}$ as 1 or 0 (i.e., all users are static), we can obtain a special case of problem \textbf{P2}, which has the same structure as a  NP-hard problem
formulated in \cite{niki2012femto}. Since \textbf{P2} is a special case of \textbf{P1}, problem \textbf{P1} is NP-hard.
As a consequence, it is impossible to find its global optimal solution within polynomial time. By using similar way of proof  in \cite{niki2012femto}, it is not hard to prove that \textbf{P1} is equivalent to maximizing a submodular function over matroid constraints. Thus, we can resort to greedy algorithm, which is commonly used to provide a solution achieving at least $\frac{1}{2}$ of the optimal value for such type of problem \cite{nemhauser1978analysis}.\footnote{The best algorithm with polynomial time complexity for such problem can achieve $(1-\frac{1}{e})$ optimality guarantee, which is based on continuous greedy process and pipage rounding techniques \cite{calinescu2011maximizing}. However, when $K=100$ and $F=3000$ in the considered setting as detailed later, its complexity is $O((FK)^8) = O(6.5\times10^{43})$, which is too complex for our problem.}

The greedy algorithm starts with zero elements for the caching matrix, i.e., $\textbf{C}= (0)^{K \times F}$. In each step, the value of one element in \textbf{C} is changed from zero to one with the maximal incremental caching gain defined as
\begin{equation}
\label{margin_f}
\begin{aligned} \textstyle
& \textstyle  v_{\bf C}(m,f) = p_{\rm off} ({{\bf {Q},{w},A,C}|_{c_{m,f}=1}}) - p_{\rm off} ({\bf {Q},{w},A,C})\\
& \textstyle \stackrel{(a)}{=}\sum_{k=1}^{K} w_k q_{f|k} \left( p^{\rm d}_{k,f} \left({\bf A,}{\bf C}|_{c_{m,f}=1}\right)-p^{\rm d}_{k,f}\left(\bf A,C\right)  \right),
\end{aligned}
\end{equation}
where $(a)$ follows by substituting \eqref{p_off}, $\bf C$ is the caching matrix at previous step, and ${\bf C}|_{c_{m,f}=1}$ is the matrix by letting $c_{m,f}=1$ in $\bf C$. The algorithm is summarized in Algorithm \ref{greedy_algo}.

\begin{algorithm}[!htb]
	\caption{ Greedy Algorithm}
	\label{greedy_algo}
	\begin{algorithmic}[1] 
		\REQUIRE ~  
		 $\bf A$;
		 $\bf {w}$;
		 $\bf {Q}$;
		Initialize:
		Caching matrix $\textbf{C}= (0)^{K \times F}$;
		Files not cached at the $m$th user $\mathcal{\bar C}_m \leftarrow \{{\rm f}_1,{\rm f}_2,...,{\rm f}_F\}$;
		Users with residual storage space $\mathcal{U}_0 \leftarrow \{{\rm u}_1,{\rm u}_2,...,{\rm u}_K\}$;
		
		\WHILE{$\mathcal{U}_0 \neq \emptyset$}
		\label{for_1}
		\STATE $[m^*,f^*] =\arg\max_{{\rm u}_m\in \mathcal{U}_0, {\rm f}_f\in\mathcal{\bar C}_m  } v_{\bf C}(m,f)$;
		\label{margin_v}
		 ${\bf C} = {\bf C}|_{c_{m^*,f^*}= 1} $;	 	
		 $\mathcal{\bar C}_{m^*} \leftarrow \mathcal{\bar C}_{m^*}\setminus {\rm f}_{f^*} $;
		\IF{$|\mathcal{\bar C}_{m^*}| = F - M$}
		\STATE $\mathcal{U}_0 \leftarrow \mathcal{U}_0 \setminus {\rm u}_{m^*} $;
		\ENDIF
		\ENDWHILE
		\STATE $\textbf{C}^* = {\bf C}$;
		\ENSURE Caching matrix $\textbf{C}^*$.
	\end{algorithmic}
\end{algorithm}

The loops in step \ref{for_1} of Algorithm \ref{greedy_algo} take  $KM$ iterations, because there are totally $KM$ files that are possible to be cached at all users. The step \ref{margin_v} for finding the element in ${\bf C}$ that introduces the highest incremental caching gain takes at most $KF$ iterations. For each time of computing $v_{\bf C}(m,f)$ in \eqref{margin_f}, the time complexity is $O(K^2)$, and thus computing all $v_{\bf C}(m,f)$ is $O(K^3F)$. Hence the total time complexity for Algorithm \ref{greedy_algo} is $O(KM(KF+K^3F)) = O(K^2FM(K^2+1))$, which is high especially when the numbers of users $K$ and files $F$ are large.


\subsection{A Low Complexity Algorithm with 1/2 Optimality Guarantee}
Since the greedy algorithm is with high time complexity, finding a low-complexity algorithm is worthwhile for practice use. In what follows, we  propose an alternating optimization algorithm, which improves the offloading gain at every iteration and converges to a local optimal solution.

To be specific, we fix the caching policy at users ${\bf c}_{m}(m\neq k', 1 \leq m \leq K)$ and optimize ${\bf c}_{k'}$. Then, from problem $\textbf{P1}$ we obtain the optimization problem with respect to ${\bf c}_{k'}$ as
\begin{equation}
\label{equ.opt6}
\begin{aligned}\textstyle
\textbf{P1}': \quad&   \textstyle  \max_{c_{k',f}}\,\, &&  \textstyle f_{\rm off}({\bf c}_{k'}) = \sum_{k=1}^{K} w_k \sum_{f=1}^{F} q_{f|k}\left(1-\prod_{m=1,m\neq k'}^{K} (1- a_{k,m}c_{m,f}) (1- a_{k,k'}c_{k',f}) \right)  \\
&  \textstyle  s.t.  && \textstyle \sum_{f=1}^{F} c_{k',f} \leq M, c_{k',f} \in \{0,1\}, 1 \leq f \leq F.
\end{aligned}
\end{equation}


\begin{proposition}
	\label{proposition3}
	$\textbf{P1}'$ can be solved with polynomial time complexity $O(F(K^2+M))$.
\end{proposition}
\begin{IEEEproof}
	See Appendix \ref{a:3}.
\end{IEEEproof}

Based on the proof of Proposition \ref{proposition3}, we propose an algorithm to iteratively solve problem $\textbf{P1}'$ by changing $k'$ from $1$ to $K$ until convergence. The algorithm starts with a given initial value of $\textbf{C}$.  In every iteration, by fixing $\textbf{c}_m  (m \neq k', 1 \leq m \leq K)$, it respectively computes the offloading gain introduced by caching the $f$th file at the $k'$th user
\begin{equation}
\label{b_k_f} \textstyle
b_{k',f} = \sum_{k=1}^{K} w_k  q_{f|k} a_{k,k'}\prod_{m=1,m\neq k'}^{K} (1- a_{k,m}c_{m,f}), 1 \leq f\leq F, 1 \leq k' \leq K.
\end{equation}
Then, the algorithm finds the file indices with the maximal $M$ values of $b_{k',f}$ to constitute a set $\mathcal{I}_{k'}$, and obtain $\textbf{c}^*_{k'}$ as
\begin{equation}
\label{equ.res.algo}
c^*_{k',f} =\left\{
\begin{aligned}\textstyle
1 & \textstyle , & \textstyle f \in \mathcal{I}_{k'} \\
0 &\textstyle , & \textstyle f \notin \mathcal{I}_{k'}
\end{aligned}.
\right.
\end{equation}
The detailed algorithm is presented in Algorithm \ref{local_opt_algo}. The loops in step \ref{for_optimal} take  $K$ iterations. Step \ref{opt_c_k} is with complexity $O(F(K^2+M))$ according to Proposition \ref{proposition3}. Hence the total time complexity of Algorithm \ref{local_opt_algo} is $O(t_{A2}KF(K^2+M))$, where $t_{A2}$ is the number of iterations for step \ref{t_iteration}.

\begin{algorithm}[!htb]
	\caption{A Low Complexity Algorithm}
	\label{local_opt_algo}
	\begin{algorithmic}[1] 
		\REQUIRE ~  
		$\bf A$;
		$\bf {w}$;
		$\bf {Q}$;
		Initialize:
		Random caching  $\textbf{c}^{(0)}_{m} (1\leq m \leq K)$, $t \leftarrow 1$;
		\REPEAT
		\label{t_iteration}
		\FOR{$k'=1,2,...,K $}
		\label{for_optimal}
		\STATE Based on $\textbf{c}^{(t-1)}_{m} (m\neq k')$, compute $b_{k',f}$ by \eqref{b_k_f}, constitute $\mathcal{I}_{k'}$ and obtain $\textbf{c}^*_{k'}$ by \eqref{equ.res.algo}.  \\
		\label{opt_c_k}
		\STATE $\textbf{c}^{(t)}_{k'} = \textbf{c}^*_{k'}$;	 		
		\ENDFOR
		\UNTIL We obtain the converged result $(\textbf{c}^{(t)}_m = \textbf{c}^{(t-1)}_m, 1\leq m \leq K)$
				\ENSURE Caching matrix $\textbf{c}^{(t)}_m$.
	\end{algorithmic}
\end{algorithm}

\begin{proposition}
	\label{proposition4}
	Algorithm \ref{local_opt_algo} monotonically increases the objective function of problem $\textbf{P1}$ and finally converges to achieve at least $1/2$ optimality.
\end{proposition}
\begin{IEEEproof}
	See Appendix \ref{a:4}.
\end{IEEEproof}


It is noteworthy that Algorithm \ref{greedy_algo} and Algorithm \ref{local_opt_algo} can also solve $\textbf{P2}$ by letting  $q_{f|k} = {p}_f, \forall k,f$ in $\bf {Q}$. Solutions for \textbf{P1} and \textbf{P2} obtained with Algorithm \ref{greedy_algo} are called ${\textbf{S1}}-{A1}$ and $\textbf{S2}-{A1}$, and solutions using Algorithm \ref{local_opt_algo} for \textbf{P1} and \textbf{P2} are called ${\textbf{S1}}-{A2}$ and $\textbf{S2}-{A2}$, respectively.

\section{Learning User Preference and Active Level}
\label{sec:predict}
In this section, we first use pLSA to model content request behavior of an individual user. We then learn the model parameters by maximizing likelihood function, either without the pLSA model for comparison or with the model using the EM algorithm, which is efficient for ML parameter estimation with latent variables \cite{demp1977maximum}.
Finally, we present a prior knowledge based algorithm to learn user preference.

\subsection{Modeling Individual User Behavior in Requesting Contents}
\label{sec.predict.plsa}
To characterize the request behavior of a user, pLSA associates each request with a topic, which may be unobservable but can be intuitively interpreted as {comedy, adventure}, etc.

By introducing latent topic set $\mathcal{Z} = \{{\rm z}_1,{\rm z}_2,...,{\rm z}_{Z} \}$ with cardinality  $|\mathcal{Z}|=Z$, pLSA associates each topic ${\rm z}_j \in \mathcal{Z}$ with each possible user request, i.e., ${\rm u}_k \in \mathcal{U}$ requests  ${\rm f}_f \in \mathcal{F}$.  Specifically, the request of each user can be modeled as the following steps with three model parameters:
\begin{itemize}
	\item A request is sent by ${\rm u}_k$ with probability $P({\rm u}_k)$ (i.e., active level),
	\item ${\rm u}_k$ chooses a topic ${\rm z}_j$ with probability $P({\rm z}_j|{\rm u}_k)$ (i.e., topic preference, $\sum_{j=1}^{Z} P({\rm z}_j|{\rm u}_k) = 1$),
	\item ${\rm u}_k$ requests  ${\rm f}_f$ in topic ${\rm z}_j$ with probability $P({\rm f}_f|{\rm z}_j)$, $\sum_{f=1}^{F} P({\rm f}_f |{\rm z}_j) = 1$, where {\em conditional independence} assumption is used. In particular, conditioned on a request being sent by  ${\rm u}_k$ who chooses topic ${\rm z}_j$, ${\rm u}_k$ chooses ${\rm f}_f$ with probability $P({\rm f}_f|{\rm z}_j,{\rm u}_k ) = P({\rm f}_f|{\rm z}_j)$, i.e., $P({\rm f}_f |{\rm u}_k ) = \sum_{{\rm z}_j \in {\mathcal{Z}}}P({\rm f}_f |{\rm z}_j) P({\rm z}_j|{\rm u}_k)$. In other words, no matter which user sends a request and selects topic  ${\rm z}_j$, the user will request ${\rm f}_f$ with probability $P({\rm f}_f|{\rm z}_j)$.
\end{itemize}

Then, the joint probability that ${\rm u}_k$ requests ${\rm f}_f$  can be expressed as
\begin{equation}
	\label{joint_p_u_f}
	\begin{aligned}\textstyle
	P({\rm u}_k  , {\rm f}_f )&  \textstyle = P({\rm u}_k  )P({\rm f}_f |{\rm u}_k )  = P({\rm u}_k ) \sum_{{\rm z}_j \in {\mathcal{Z}}}P({\rm f}_f |{\rm z}_j) P({\rm z}_j|{\rm u}_k).
	\end{aligned}
\end{equation}

\subsection{Learning Individual User Behavior in Requesting Contents}
   According to maximal likelihood (ML) principle, we can learn $ P({\rm u}_k ), P({\rm f}_f |{\rm z}_j)$ and $P({\rm z}_j |{\rm u}_k)$ with requests history $n_{k , f}$ by maximizing the following log-likelihood function \cite{hofmann2001unsupervised}
\begin{equation}
\label{likelihood}
\begin{aligned} \textstyle
\mathcal{L} = \sum_{i} \log P({\rm u}_{i_{\rm u}},{\rm f}_{i_{\rm f}})= \underbrace{\sum_{{\rm u}_k \in \mathcal{U}} \sum_{{\rm f}_f \in \mathcal{F}} n_{k , f}\log P({\rm u}_k , {\rm f}_f )}_{(a)} =  \underbrace{\sum_{{\rm u}_k \in \mathcal{U}} \sum_{{\rm f}_f \in \mathcal{F}}n_{k , f}\log  P({\rm u}_k ) \sum_{{\rm z}_j \in {\mathcal{Z}}}P({\rm f}_f |{\rm z}_j) P({\rm z}_j|{\rm u}_k)}_{(b)},
\end{aligned}
\end{equation}
where the $i$th sample corresponding to the event that the $i_{\rm u}$th user requests the $i_{\rm f}$th file, and in (b) the pLSA model is applied.

\subsubsection{ML algorithm without pLSA model} By maximizing the log likelihood function in (a) of \eqref{likelihood} without the pLSA model, it is not hard to obtain that
\begin{equation}
\label{closed_estimate}\textstyle
\hat P({\rm u}_k , {\rm f}_f )= \frac{n_{k,f}}{\sum_{k'=1}^{K}\sum_{f=1}^{F} n_{{k'}, f}}.
\end{equation}

Then, the active level and user preference can be learned as $\hat{w}_k = \hat P({\rm u}_k) = \sum_{f=1}^{F}\hat P({\rm u}_k , {\rm f}_f )$ and $\hat q_{f|k}= \frac{\hat P({\rm u}_k , {\rm f}_f )}{\hat P({\rm u}_k)} = \frac{\hat P({\rm u}_k , {\rm f}_f )}{\sum_{f=1}^{F}\hat P({\rm u}_k , {\rm f}_f )}$, respectively. This algorithm is actually a simple frequency-count prediction, which can serve as a baseline for learning active level and user preference.

{\bf Remark 4:} If we directly predict $w_k$ and $q_{f|k}$ using \eqref{closed_estimate}, the number of parameters to estimate is $KF$. By using the pLSA as in (b) of \eqref{likelihood}, the number of parameters is reduced from $KF$ to $K+KZ+ZF = Z(K+F)+K$, where  $K$ parameters are for learning active level, $KZ$ parameters are for topic preference, and $ZF$ parameters are for $P({\rm f}_f|{\rm z}_j)$. With less number of parameters to estimate, a learning algorithm can converge more quickly.

\subsubsection{ML algorithm with pLSA model} To maximize the log-likelihood function in (b) of \eqref{likelihood}, we first rewrite the function as
\begin{equation}
\label{likelihood_EM}\textstyle
\mathcal{L} = \underbrace{\sum_{{\rm u}_k \in \mathcal{U}}   n_k \log  P({\rm u}_k )}_{(a)} +  \underbrace{\sum_{{\rm u}_k \in \mathcal{U}} \sum_{{\rm f}_f \in \mathcal{F}} n_{k , f} \log  \sum_{{\rm z}_j \in {\mathcal{Z}}}P({\rm f}_f |{\rm z}_j) P({\rm z}_j|{\rm u}_k)}_{(b)},
\end{equation}
where $n_k =  \sum_{{\rm f}_f \in \mathcal{F}} n_{k , f}$. It is not hard to see that the terms in (a) and (b) can be independently maximized. The active level  of ${\rm u}_k$ can be learned by maximizing term (a) in \eqref{likelihood_EM} as
\begin{equation}
\label{p_u_k}\textstyle
	\hat{w}_k = \hat P({\rm u}_k ) = \frac{n_k}{\sum_{k'=1}^{K}\sum_{f=1}^{F} n_{k' , f}},
\end{equation}
which is the same as that obtained from \eqref{closed_estimate}. The other two model parameters $P({\rm f}_f |{\rm z}_j)$ and  $P({\rm z}_j|{\rm u}_k)$ can be learned by maximizing term (b) in \eqref{likelihood_EM} using the EM algorithm as follows \cite{hofmann2001unsupervised}.

Starting from randomly generated initial values for the model parameters $P({\rm z}_j|{\rm u}_k)$ and $P({\rm f}_f |{\rm z}_j)$,  $1 \leq j \leq Z$, $1 \leq f \leq F$ and $1 \leq k \leq K$, the EM algorithm alternates two steps: expectation (E) step and maximization (M) step.
In the E-step, the posterior probabilities are computed for latent variable ${\rm z}_j$ with current estimation of the parameters as
\begin{equation}
\label{E_step}\textstyle
\hat P({{\rm z}_j|{\rm u}_k,{\rm f}_f})= \frac{\hat  P({\rm z}_j|{\rm u}_k)\hat P({\rm f}_f |{\rm z}_j)}{\sum_{{{\rm z}_{j'}} \in \mathcal{Z}} \hat P({\rm z}_{j'}|{\rm u}_k) \hat P({\rm f}_f |{\rm z}_{j'})},
\end{equation}
which is the probability that ${\rm f}_f$ requested by  ${\rm u}_k$ belongs to topic ${\rm z}_j$.
In the M-step, given $\hat P({{\rm z}_j|{\rm u}_k,{\rm f}_f})$ computed by previous E-step, the parameters are updated  as
\begin{equation}\label{M-3}
	\textstyle  \hat P({\rm f}_f|{\rm z}_j) =\frac{ \sum_{{\rm u}_k \in \mathcal{U}}  n_{k , f} \hat P({\rm z}_j|{\rm u}_k,{\rm f}_f)    }
	{   \sum_{{\rm u}_k \in \mathcal{U}} \sum_{{\rm f}_{f'} \in \mathcal{F}}   n_{k , f'}  \hat P({\rm z}_j|{\rm u}_k,{\rm f}_{f'}) }, \quad {\rm and} \quad  \hat P({\rm z}_j|{\rm u}_k) = \frac{  \sum_{{\rm f}_f \in \mathcal{F}} n_{k , f}  \hat P({\rm z}_j|{\rm u}_k,{\rm f}_f)   }
	{n_k }.
\end{equation}

By alternating \eqref{E_step} with \eqref{M-3}, the EM algorithm converges to a local maximum of log-likelihood function.
Then, the preference of the $k$th user for the $f$th file can be learned as \begin{equation}
\hat q_{f|k}= \hat P({\rm f}_f|{\rm u}_k) = \sum_{{\rm z}_j \in {\mathcal{Z}}}\hat P({\rm f}_f |{\rm z}_j) \hat P({\rm z}_j|{\rm u}_k).
\end{equation}

\vspace{-4mm}\subsubsection{Prior Knowledge Based Algorithm to Learn User Preference}\label{subsecpriori} Video files in real world website always have topic information, e.g., movies are labeled with {\em comedy}, {\em drama} and so on.

Intuitively, the topic preference and active level of a user change slowly over time, and hence can be regarded as invariant. This will be validated later by real dataset. Thanks to the pLSA model, such intuition naturally yields a {\em prior knowledge based algorithm} to learn user preference by exploiting the active level and topic preference of a user learned previously during a much longer time than learning user preference, with the help of the topic information. This algorithm can be regarded as a parameter-transfer
approach \cite{TL2010}. While the active level $P({\rm u}_k)$ and topic preference $P({\rm z}_j|{\rm u}_k)$ can never be learned perfectly, we assume that they are known in order to show the potential of such transfer learning.
Then, the user preference can be learned by only estimating $P({\rm f}_f|{\rm z}_j)$, which can be obtained similarly as in \eqref{M-3},
\begin{equation}
	\label{M-2-new} \textstyle
	 \hat P({\rm f}_f|{\rm z}_j) = \left\{
	 \begin{aligned}
	 & \textstyle\frac{ \sum_{{\rm u}_k \in \mathcal{U}}  n_{k , f} \hat P({\rm z}_j|{\rm u}_k,{\rm f}_f)    }
	 {   \sum_{{\rm u}_k \in \mathcal{U}} \sum_{{\rm f}_{f'} \in \mathcal{F}}   n_{k , f'} \hat P({\rm z}_j|{\rm u}_k,{\rm f}_{f'}) },& \textstyle \ & \textstyle {\rm f}_f \in \mathcal{F}_j \\
	 &\textstyle 0 ,& \textstyle\ & \textstyle{\rm f}_f \notin \mathcal{F}_j
	 \end{aligned},
	 \right.
\end{equation}
where $\mathcal{F}_j$ is the set of files associated with topic ${\rm z}_j (1\leq j \leq Z)$, which is available on the video website. For instance, the movie {\em Forrest Gump} is associated with topics {\em comedy, romance} and {\em war} on the MovieLens.  The detailed  algorithm  is presented in Algorithm \ref{prior_knowledge}. Step 2 takes  $KFZ$ times computation of posterior probability by \eqref{E_step}, where each computation is with complexity $O(Z-1)$ and thus totally at most $O(KFZ(Z-1))$. Step 3 computes \eqref{M-2-new} with $ZF$ times, each is with complexity $O(K(F+1))$. It is not hard to see that step 4 is with complexity $O(KFZ)$. Hence the total time complexity for Algorithm \ref{prior_knowledge} is $O(t_{A3}KFZ(Z+F+1))$, where $t_{A3}$ is the number of iterations for step \ref{t_iteration}.

\begin{algorithm}[!h]
	\caption{Learning user preference with prior knowledge.}
	\label{prior_knowledge}
	\begin{algorithmic}[1] 
		\REQUIRE ~
		 ${\bf N} $; $Z$; $\mathcal{F}_j,  1 \leq j\leq Z$;  $\hat P({\rm z}_j |{\rm u}_k)$;
		Stop condition $0 <\epsilon < 1$;  \\
		Initialize:
		$\hat P^{(0)}({\rm f}_f |{\rm z}_j)$; Step $i \leftarrow 1$; Difference $\Delta \leftarrow \infty$;
		Log likelihood $\mathcal{L}(0)\leftarrow0$;
		
		\WHILE{$\Delta > \epsilon$}
		\STATE Using  $\hat P({\rm z}_j |{\rm u}_k)$ and $\hat P^{(i-1)}({\rm f}_f |{\rm z}_j)$ to compute $\hat P^{(i)}({\rm z}_j|{\rm u}_k,{\rm f}_f)$ by  \eqref{E_step};
		\STATE Using $\hat P^{(i)}({\rm z}_j|{\rm u}_k,{\rm f}_f)$ and $\mathcal{F}_j$ to compute $\hat P^{(i)}({\rm f}_f |{\rm z}_j)$ by \eqref{M-2-new};
		\STATE Compute log likelihood $\mathcal{L}(i)$ with $\hat P({\rm z}_j |{\rm u}_k) $ and $\hat P^{(i)}({\rm f}_f |{\rm z}_j)$ using term (b) in \eqref{likelihood_EM};
		\STATE $\Delta = |\mathcal{L}(i) - \mathcal{L}(i-1)  | $;	 $i \leftarrow i + 1$;
		\ENDWHILE
		\STATE $\hat{q}_{f|k} \leftarrow   \sum_{{\rm z}_j \in {\mathcal{Z}}} \hat P({\rm f}_f |{\rm z}_j) \hat P({\rm z}_j|{\rm u}_k)  $ to compute $\bf \hat{Q}$;
		\ENSURE $\bf \hat{Q}$.
	\end{algorithmic}
\end{algorithm}

\section{User Request Behavior Analysis with MovieLens Dataset}\label{dataset}
The gain from caching highly depends on the user behavior in requesting contents,
both collectively and individually. In this section, we first use a real dataset to analyze the connection between file catalog size and number of users in a region, as well as the active level, topic preference of each user and user preference, and validate the intuition in Section \ref{subsecpriori}. Then, we validate the user preference model provided in Section II.

\vspace{-2mm}\subsection{Statistical Results of User Demands}
\label{topic_info}
We use the {\em MovieLens 1M Dataset} \cite{harper2016movielens} to reflect the request behavior, where {\em MovieLens} is a website that recommends movies for its users operated by {\em GroupLens} lab at the {\em University of Minnesota}.
This dataset contains $1000209$ ratings for $3952$ movies provided by $6026$ users  from the year of 2000 to 2003. Each sample of the dataset consists of user identity (ID), movie ID, rating  and timestamp. Because users typically give ratings only after watching, we translate  the rating record into the request record, i.e., when a user gives rating for a movie, we set the movie as requested by once. Except for the ratings, MovieLens also provides topic information of movies. Every movie is associated with one, two or more topics from $18$ topics, which include {\em action, adventure, animation, children's, comedy, etc.} genre and are denoted as ${\rm z}_1$, $...$, ${\rm z}_{18}$. For instance, {\em Forrest Gump} is associated with topics {\em comedy (${\rm z}_5$), romance (${\rm z}_{14}$)} and {\em war (${\rm z}_{17}$)}. From the topic information provided by MovieLens, we can see that if the $f$th movie is not associated with the $j$th topic, users who select $j$th topic will not request the $f$th file, i.e., we can set $P({\rm f}_f|{\rm z}_j) = 0$  in \eqref{M-2-new}.

To analyze temporal evolution of user behavior, we sort all the $3952$ movies according to their released date in ascendant order  and then divide them into two subsets $\mathcal{F}_1$ and $\mathcal{F}_2$, where the file request matrices on $\mathcal{F}_1$ and $\mathcal{F}_2$ are $\textbf{N}_1 \in \mathbb{R}^{6040\times 1976 }$ and $\textbf{N}_2 \in \mathbb{R}^{6040\times 1976 }$, respectively. $\textbf{N}_1$ can reflect user behavior on previously released file subset $\mathcal{F}_1$, and $\textbf{N}_2$ can reflect user behavior on subsequently released file subset $\mathcal{F}_2$.
Specifically, we analyze the following statistical results:
\begin{itemize}
	\item {\em File catalog size:} To reflect the randomness of the users in
sending requests, it is the average number of files requested by a given number of randomly chosen users, which is obtained from $\textbf{N} = [\textbf{N}_1 \  \textbf{N}_2 ]$ and the average is taken over users.

	\item {\em Active level:} $P_1({\rm u}_k)$ and $P_2({\rm u}_k)$ are computed using \eqref{p_u_k} with $\textbf{N}_1 $ and $\textbf{N}_2$, respectively.
	
	\item {\em Topic preference:} Denote the topic preference of the $k$th user estimated on subsets $\mathcal{F}_1$ and $\mathcal{F}_2$ as ${\textbf{p}}_1(\mathcal{Z}|{\rm u}_k) = [P_1(\rm{z}_1|{\rm u}_k),...,P_1(\rm{z}_Z|{\rm u}_k)]$ and ${\textbf{p}}_2(\mathcal{Z}|{\rm u}_k) = [P_2(\rm{z}_1|{\rm u}_k),...,P_2(\rm{z}_Z|{\rm u}_k)]$, respectively. $P_1(\rm{z}_j|{\rm u}_k)$, $P_2(\rm{z}_j|{\rm u}_k)$ are computed using \eqref{M-3} by EM algorithm with $\textbf{N}_1 $ and $\textbf{N}_2$, respectively. To reflect the temporal dynamic of topic preference for the $k$th user, we use the metric of cosine similarity in \eqref{cos_sim} to evaluate the similarity level as $\text{sim} ({\textbf{p}}_1(\mathcal{Z}|{\rm u}_k),{\textbf{p}}_2(\mathcal{Z}|{\rm u}_k))$.
	\item {\em User preference:}  $q_{f|k}=  \sum_{{\rm z}_j \in {\mathcal{Z}}}P({\rm f}_f |{\rm z}_j) P({\rm z}_j|{\rm u}_k)$ is obtained by EM algorithm on $\textbf{N}_1$. The result obtained from $\textbf{N}_2$ or $\textbf{N}$ is similar, and hence is omitted for conciseness.
\end{itemize}

\begin{figure}[!htb]
	\centering
	\subfigure[Connection between file catalog size and number of users]{
		\includegraphics[width=0.45\textwidth]{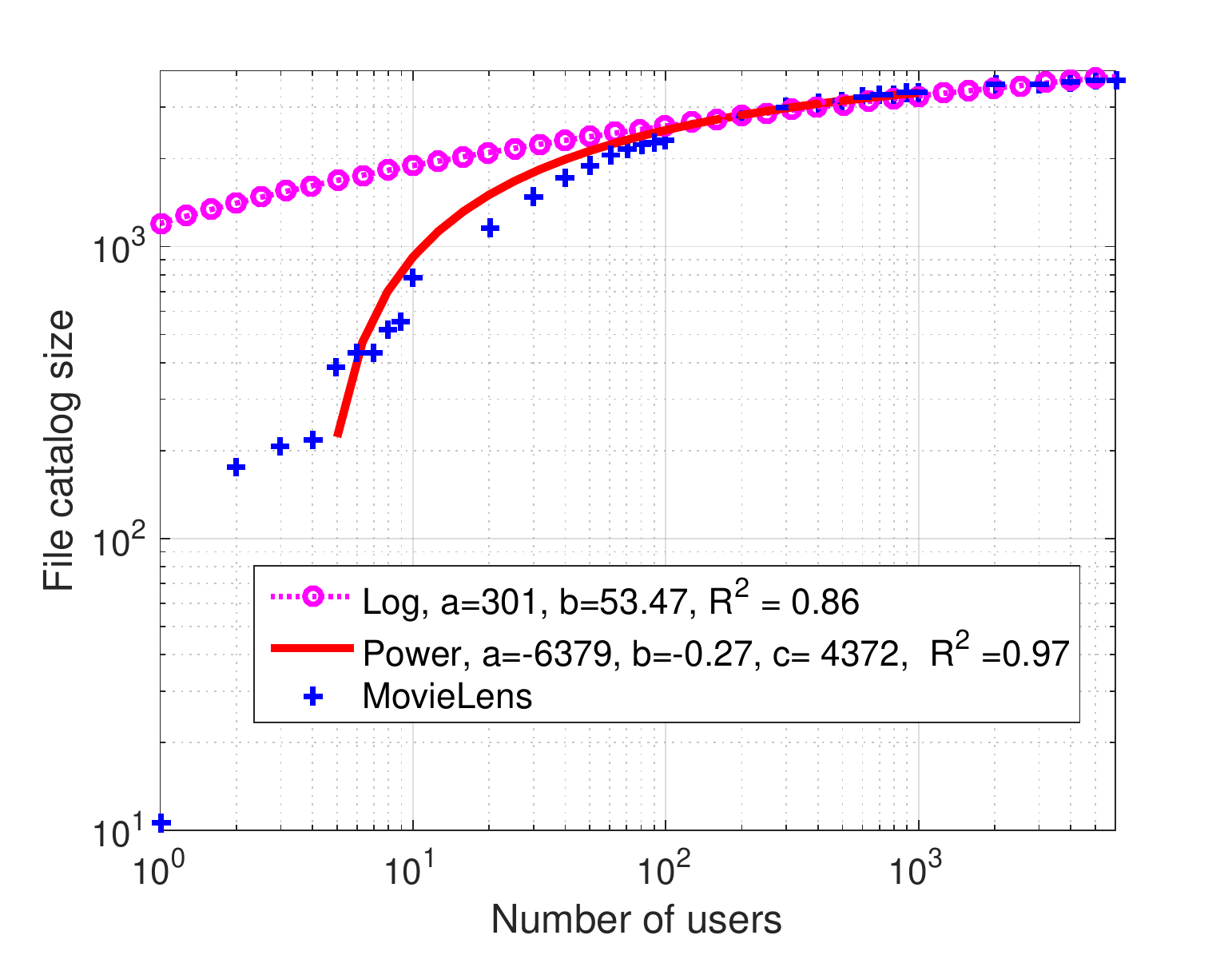}
	}
	\subfigure[Active levels of different users in descending order]{
		\includegraphics[width=0.45\textwidth]{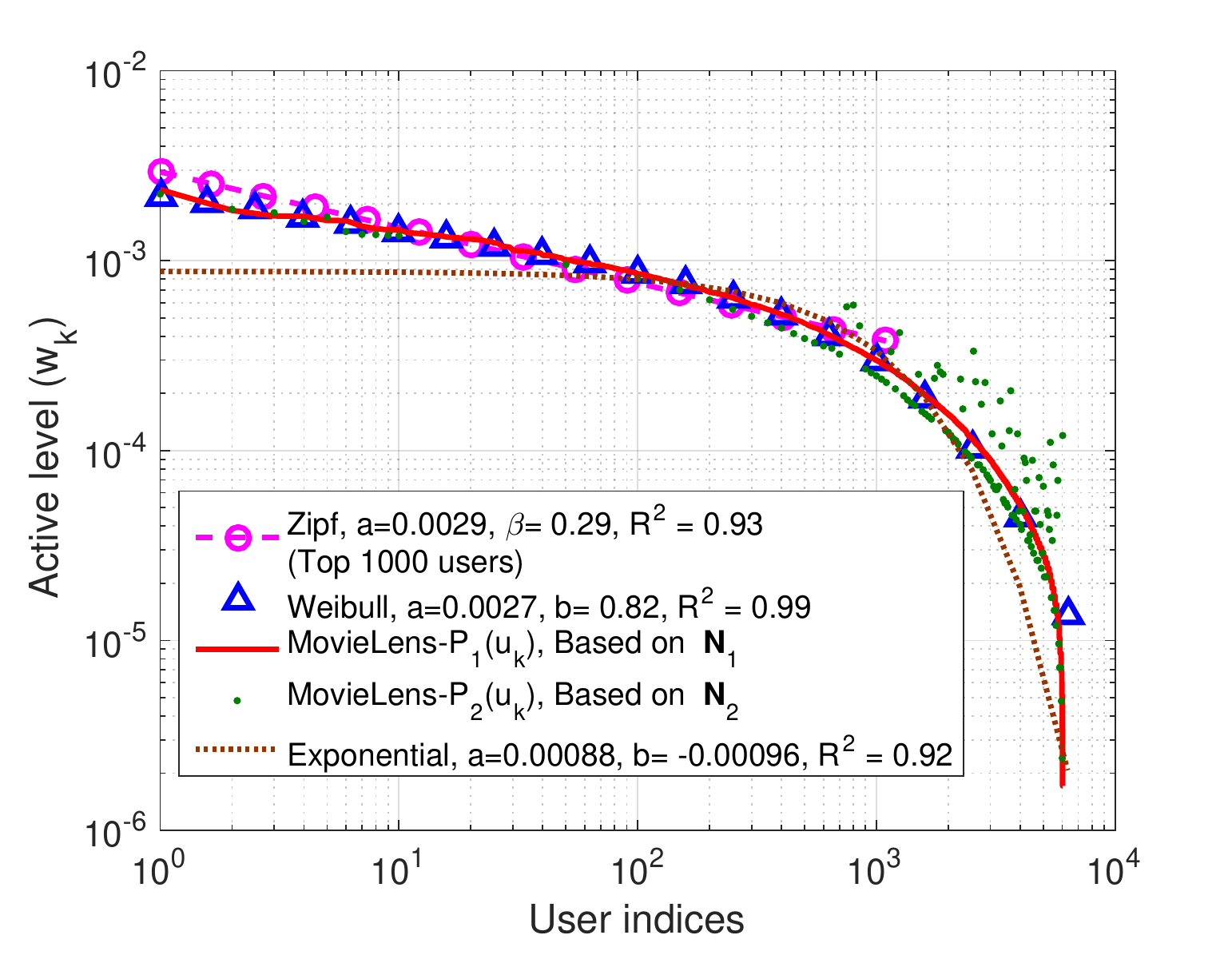}}
	\caption{File catalog size and user active level in log-log coordinates.}\label{fig.1n}
\end{figure}

In Fig. \ref{fig.1n}(a), we show the relation between file catalog size and the number of users obtained
from dataset (with legend ``MovieLens'') and the fitted curves. The curve with legend ``Log"  fits the function $f(x)= {\rm a}\log({\rm b}x)$, and the curve with legend ``Power" is with  $f(x) = {\rm a}x^{{\rm b}} + {\rm c}$. To evaluate the goodness of fit, we use the coefficient of determination (also called R-square) in linear regression, i.e.,
$R^2 = 1 - \frac{\sum_{i=1}^{S} (y_i - f(x_i))^2}{\sum_{i=1}^{S} (y_i - \bar{y})^2}$, where $S$ is the number of samples, $(x_i,y_i)$ is the $i$th sample, and $\bar{y} = \frac{\sum_{i=1}^{S} y_i}{S}$ \cite{Trivedi2002Probability}.  $R^2 \leq 1$,  and the large value of $R^2$ indicates good fitting result.  The parameters $a$, $b$ and $c$ for each function and $R^2$ are listed in the legends. We can see that the catalog size first increases quickly and then slowly, where ``Power" function fits better than ``Log" function. When the number of users is small (e.g., in a small cell), the catalog size is small, which implies that the cache hit ratio could be high with limited cache size. However, with limited number of requests due to a few associated users, the popularity is hard to predict rapidly at the small BS. When the number of users is large (e.g., in a macro cell), the catalog size increases slowly, and both fitted curves are close to the measured catalog size. In \cite{dongcaching16}, the authors suggest to use logarithm function to compute the catalog size without validation. Here, the result shows that  ``Log" is reasonable when the number of users in the considered area is large, say $K \ge 100$, at least for MovieLens dataset.

In Fig. \ref{fig.1n}(b), we show the active levels of users, where the user indices are ranked in descending order according to $P_1({\rm u}_k)$. We can see that the distribution of active levels is skewed, which indicates that majority requests are generated by a small number of users. Besides, the distribution of active levels from the two subsets of data  are similar, where the cosine similarity is $0.87$. This validates that the active level of a user changes slightly over time.  We also show the corresponding fitted distributions, where ``Weibull" is with function $f(x) = {\rm ab}x^{\rm b-1}e^{-{\rm a}x^{\rm b}} $, ``Exponential" is with function $f(x) = {\rm a}e^{{-\rm b}x}$, and ``Zipf"  is with function $f(x)= {\rm a}x^{-\rm \beta}$ (shown only for the most active $1000$ users). We can see that the curve with the real data is linear on a log-log scale for active users, but the tail (after the $1000$th user) decreases quickly. The truncate tail may come from the rating behavior of users for watched movies on MovieLens website.  Some users rarely give ratings, and some users do not continuously give ratings. As a result, the observed active levels of these users are very low. From the values of $R^2$, we can find that ``Weibull" is the best fitted distribution.
Nonetheless, the distribution of the most active 1000 users is very close to Zipf.

\begin{figure}[!htb]
	\centering
	\subfigure[Topic preferences of different users]{
		\includegraphics[width=0.45\textwidth]{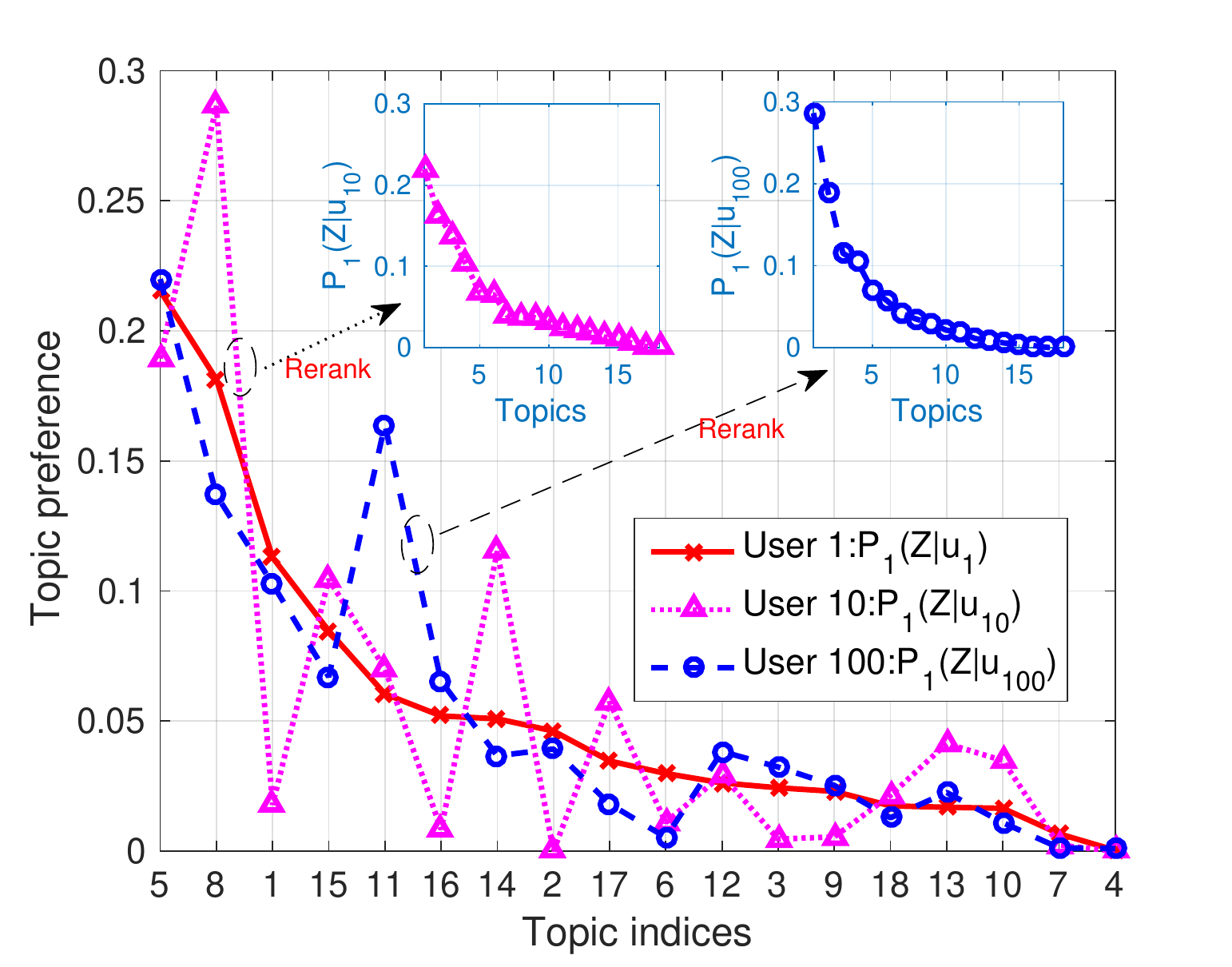}}
	\subfigure[Topic preference of the $1$st user in descending order]{
		\includegraphics[width=0.45\textwidth]{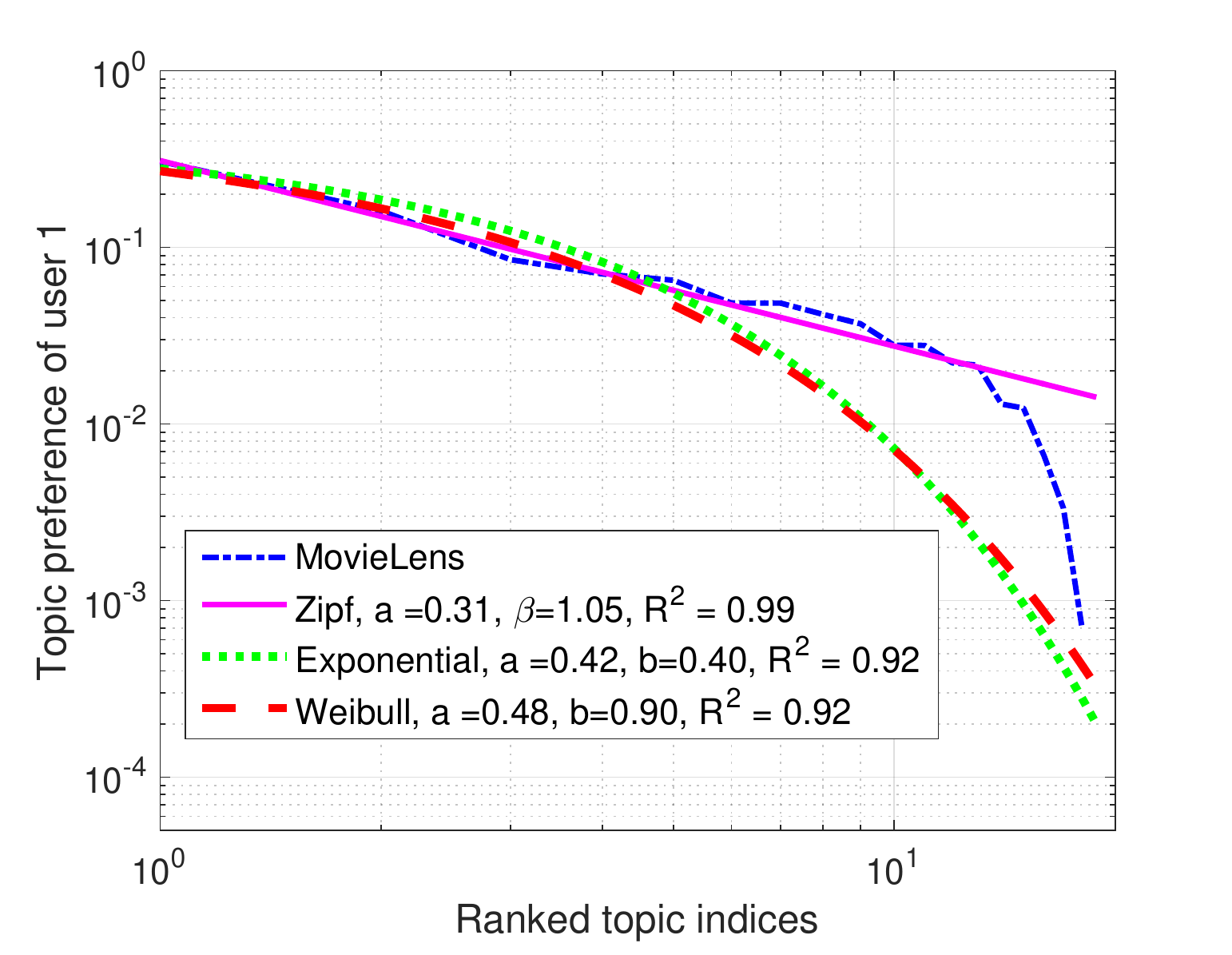}}
	\caption{Topic preferences of different users and fitted distributions.}\label{fig.2n}
\end{figure}

In Fig. \ref{fig.2n}(a), we show the topic preferences of the $1$st, $10$th and $100$th users obtained from $\mathcal{F}_1$, i.e., $  {\textbf{p}}_1(\mathcal{Z}|{\rm u}_1)$, $ {\textbf{p}}_1(\mathcal{Z}|{\rm u}_{10})$ and $ {\textbf{p}}_1(\mathcal{Z}|{\rm u}_{100})$. The results obtained from $\mathcal{F}_2$ are similar and are not shown. The labels of x-axis are ranked in descending order according to $  {\textbf{p}}_1(\mathcal{Z}|{\rm u}_1)$. The topic preferences of the $10$th and $100$th users with re-ranked x-axis according to $ {\textbf{p}}_1(\mathcal{Z}|{\rm u}_{10})$ and $ {\textbf{p}}_1(\mathcal{Z}|{\rm u}_{100})$ are also provided in the inner-figures.  We can see that topic preference of each user is skewed, which indicates that each user has  strong preferences towards specific topics. In fact, the topic preferences of all users in the dataset are skewed, which is not shown for consciousness. We can also see that topic preferences of different users differ. For example, the most favorite topic is {\em comedy} for the $1$st and $100$th user and {\em drama} for the $10$th user.

In Fig. \ref{fig.2n}(b), we show the topic preference of the $1$st user and the fitted distributions in a log-log coordinate. We can see that the best fitted distribution is a Zipf distribution with parameter $\beta = 1.05$. We have also fitted distributions of topic preferences for other users, but the curves are not provided. We observe that the best fitted distributions differ for users, where Zipf distribution is the best of $1425$ users, Weibull distribution is the best for $1899$ users, and Exponential distribution is the best for the remaining $2702$ users (but the difference in $R^2$ from Weibull distribution for these users is negligible). For the users whose best fitted distributions are Zipf distributions, the parameters of $\beta$ differ, which approximately follow a uniform distribution in $[0.5, 3]$. Yet for the most favorite several topics, Zipf distribution is always the best.


\begin{figure}[!htb]
	\centering
	\subfigure[Empirical CDF and PDF of cosine similarity between ${\textbf{p}}_1(\mathcal{Z}|{\rm u}_k)$ and ${\textbf{p}}_2(\mathcal{Z}|{\rm u}_k)$.]{
		\includegraphics[width=0.45\textwidth]{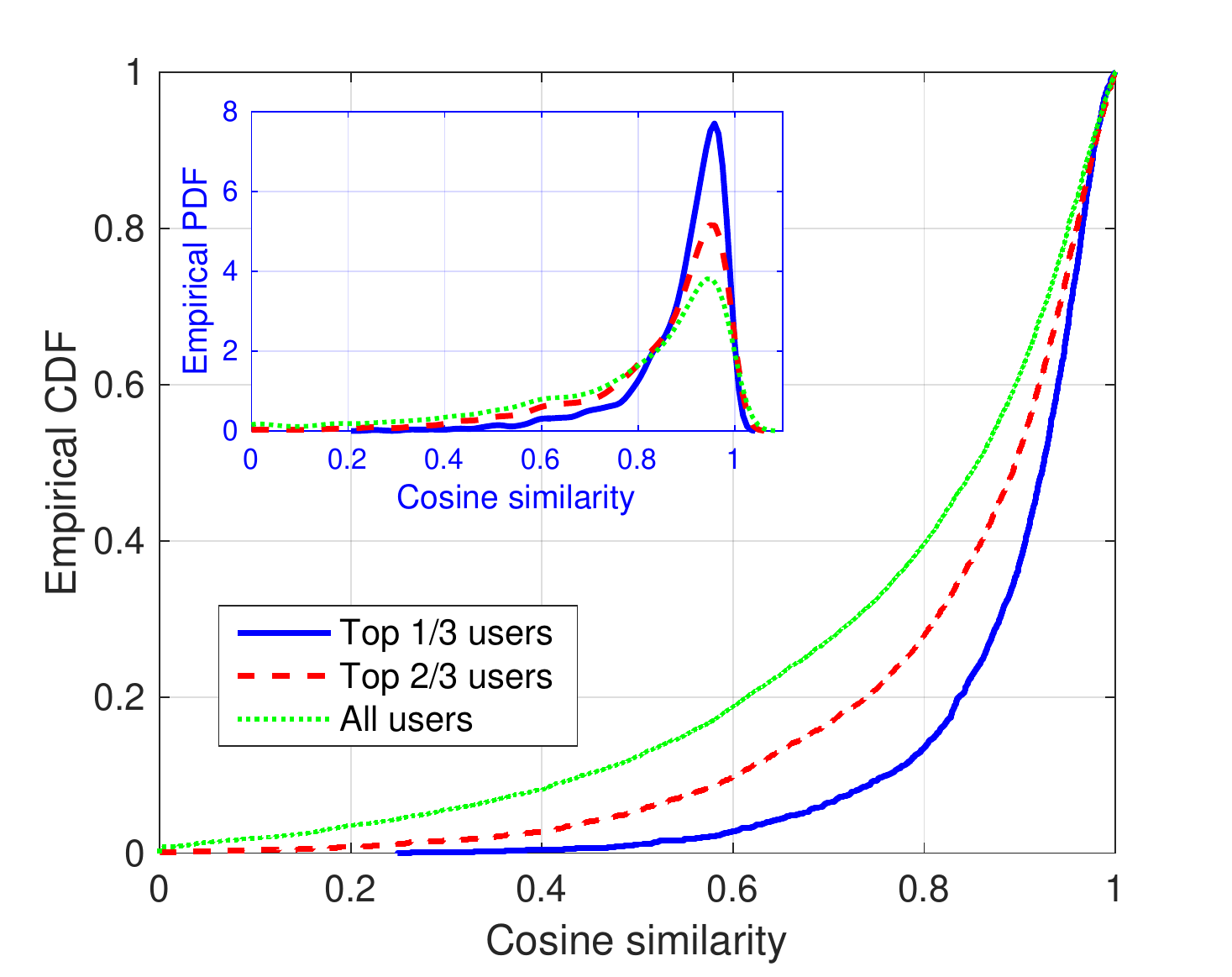}}
	\subfigure[User preference of the 1st user, where the files are ranked in descending order.]{
		\includegraphics[width=0.45\textwidth]{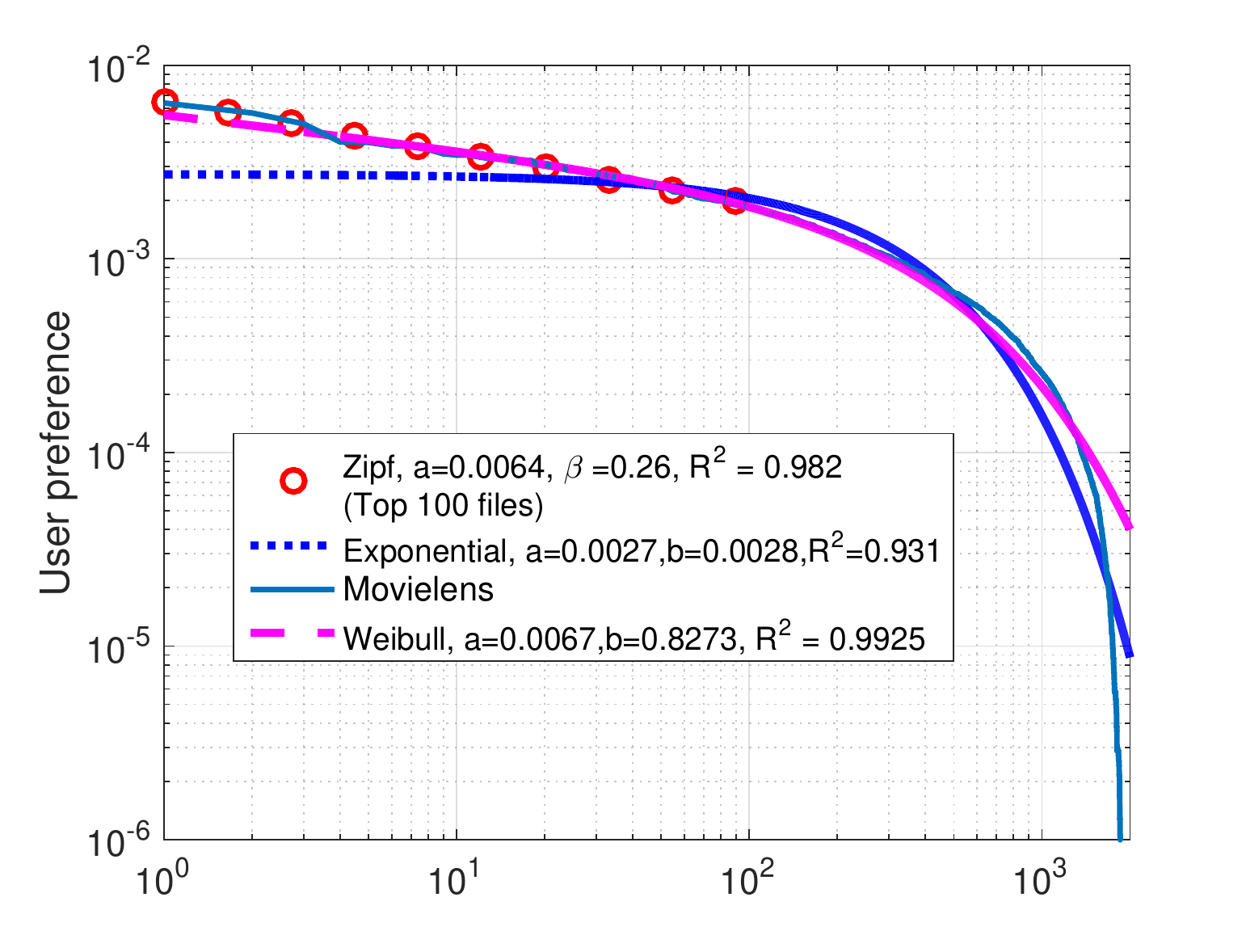}}
	\caption{Temporal dynamics of topic preferences and user preference of the 1st user.}\label{fig.3n}
\end{figure}

%


In Fig. \ref{fig.3n}(a), we show the empirical cumulative distribution function (CDF) and probability density function (PDF) of the cosine similarity between {\em topic preferences} over time of all users. We can see that $60\%$ of all users have cosine similarity larger than $0.8$, and almost $90\%$ users among the top $1/3$ active users have cosine similarity larger than $0.8$ (i.e., their topic preferences change slowly in the three years). Considering that the statistical results for active users with more requests are with high confidence level, this result indicates that topic preferences can be approximated as invariant over time.\footnote{In recommendation problem, it has been shown that user preference varies over time due to the dynamic of file catalog and the user's exploration for new items \cite{rafailidis2016modeling}. However, the topic preference variation has never been analyzed in the literature.} This validates the intuition in Section \ref{subsecpriori}.

%

In Fig. \ref{fig.3n}(b), we show user preference of the $1$st user and the fitted distributions. The user preference for the top $100$ favorite files is close to Zipf distribution (a straight line in the log-log  coordinate), and the preference for less favorite files has a truncated tail. This is because a user almost does not request the files belonging to its unfavorable topics. We have also fitted the preferences of other users (but are not shown). We find that Weibull is the best fitted distribution for all users, but the parameters and the skewness of curves  differ. The user preferences for the top favorable $100$ files of each user can be fitted with Zipf distribution, but the parameters of $\beta$ differ in a range of $[0.2, 0.8]$. In summary, the preferences of different users differ in the favorite file set, skewness, and ranking. This is not consistent with the model that all user preferences are Zipf distributions with same parameter but with different ranks as assumed in \cite{zhang2016clustered,wu2016semigradient}. Besides, we observe that the average cosine similarity of preferences among different users on dataset $\textbf{N}_1$ is as low as $ \mathbb{E}_{k,m} [ \text{sim}({\bf q}_k,{\bf q}_m) ] \approx 0.4$.\footnote{We also analyze a real video dataset of Youku in a university campus. The result shows that $ \mathbb{E}_{k,m} [ \text{sim}({\bf q}_k,{\bf q}_m) ] \approx 0.28$.}  This is mainly because the interested file sets of users are less overlapped, recalling that the topic preferences of users differ.

\subsection{Validating Synthetic User Preference Model}
\label{sec.model_UP}


Now, we validate the user preference model by comparing the results obtained from data synthesized by the generative process in Section II and those from the {\em MovieLens} dataset.

\begin{figure}[!htb]
	\centering
	\subfigure[Active levels of users in a log-log coordinate]{
		\includegraphics[width=0.475\textwidth]{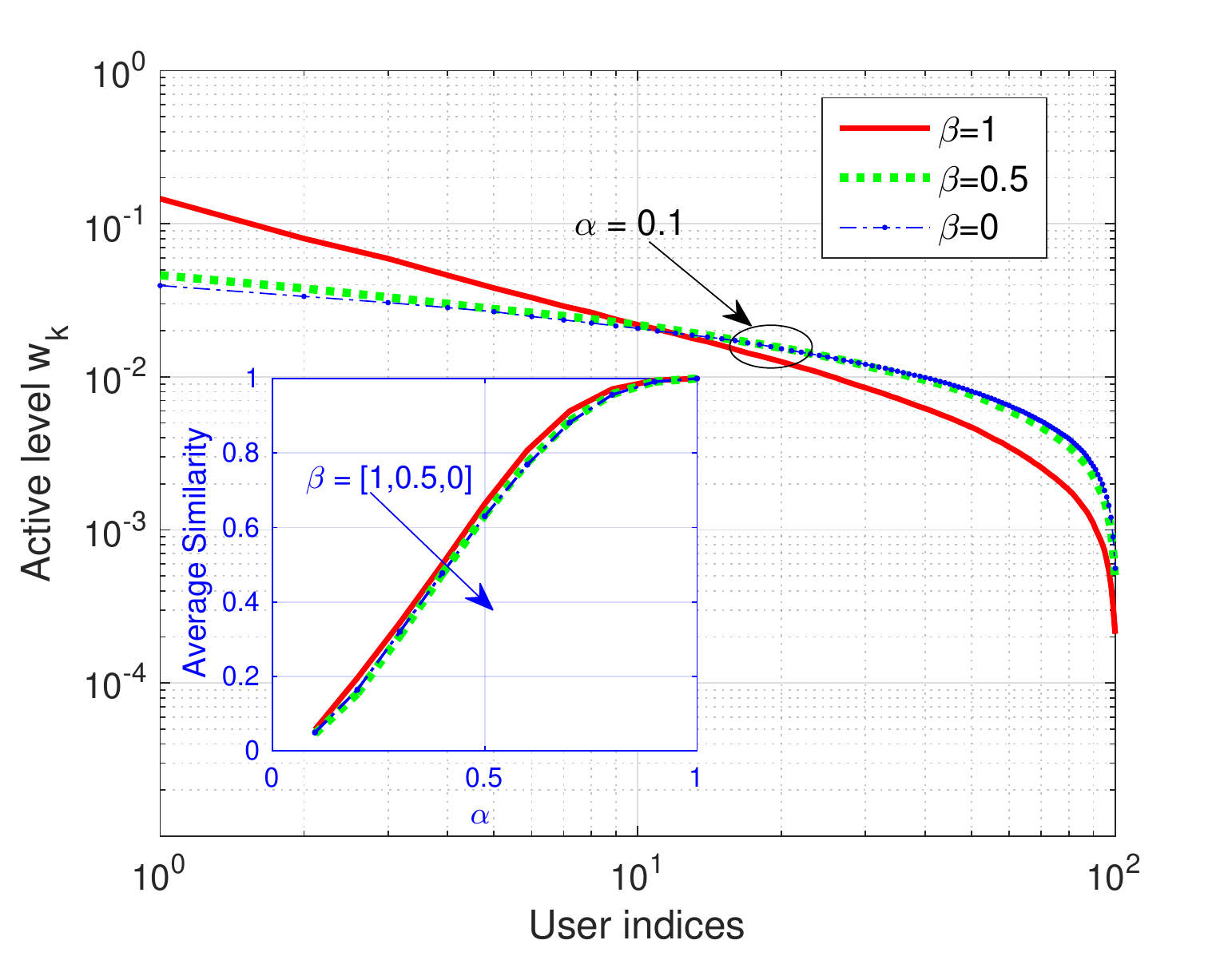}
	}
	\subfigure[Topic preferences of users, $\alpha = 0.36$ and $\beta = 0.6$.]{
		\includegraphics[width=0.475\textwidth]{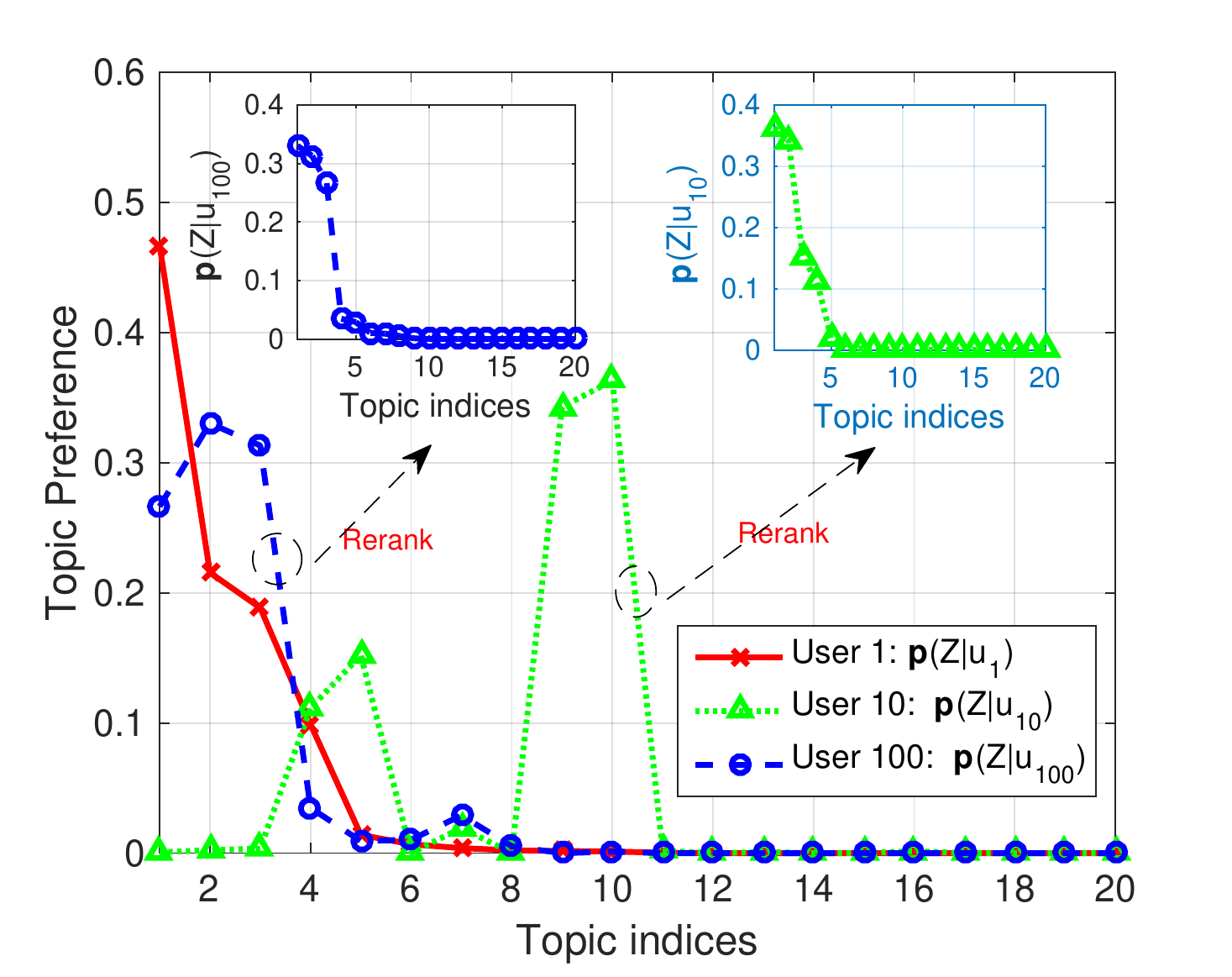}}
	\caption{Active level and topic preferences obtained from the requests generated by the synthetic method, $K=100$, $F=3000$.}\label{fig.2}
\end{figure}

In Fig. \ref{fig.2}(a), we first show the impact of parameter $\alpha$ in the kernel function. The inner-figure indicates that the synthetic
user preference model can capture different levels of similarity among user preferences by adjusting $\alpha$, while the Zip parameter $\beta$ has negligible impact on the average cosine similarity. This seems counter-intuitive, since a more skewed popularity distribution seems to imply highly correlated user preferences. However, such an intuition comes from the implicit assumption that the users send their requests with equal probability (i.e., with identical active level), which is not true in reality. From the figure we can observe that even when $\beta=1$, $\alpha$ can be as small as $0.1$. This is because few users are very active in sending file requests and have skewed user preference, who have large impact on content popularity according to \eqref{p_f}. We can see that the distributions of user active level are skewed, which agree well with the results in Fig. \ref{fig.1n}(b) obtained from the {\em MovieLens} dataset.

In Fig. \ref{fig.2}(b), we show the topic preference of the $1$st, $10$th and $100$th users. The labels of x-axis are ranked according to $  {\textbf{p}}(\mathcal{Z}|{\rm u}_1)$, as in Fig. \ref{fig.2n}(a). We can see that the topic preferences of the users are skewed, and the topic preferences of the three users are with different  distributions, which are consistent with the results in Fig. \ref{fig.2n}(a) obtained from the {\em MovieLens} dataset.


\section{Simulation Results}
\label{sec:simulation}
In this section, we demonstrate the caching gain by exploiting user preference over content popularity, either {\em perfect} or {\em predicted}. Simulation results are obtained from data generated by the synthetic model in section \ref{sec.model_UP}, which can provide ground truth of the request behavior for evaluation, and from the MovieLens dataset, which can validate the gain from real data.

We consider a square area with side length $500$ m, where $K=100$ users are  uniformly located. The collaboration distance $r_c = 30$ m. The file catalog size $F=3000$ (obtained from Fig. \ref{fig.1n}(a) for 100 users), and each user is willing to cache $M=5$ files (i.e., $1.67$ \textperthousand \ of all files).
$\alpha=0.36$ in the kernel function of the synthetic model, which corresponds to the average cosine similarity $0.4$ of the MovieLens dataset (obtained from Fig. 5(a)). The parameter of Zipf distribution is $\beta =0.6$, which is slightly smaller for a small area than that is observed at the Web proxy  as reported in \cite{gill2007youtube}.  We divide time into two-hour periods, each consisting of a peak time and an off-peak time.\footnote{Using other values as the period does not affect the learning performance of user preferences. Yet the period can not be too short, since a frequent content placement brings extra traffic load. Numerical results under the considered setup show large opportunity of using multicast for precaching. In particular, (i) only 76 files need to be pre-downloaded to users via unicast where the other 424 files can be placed to users via multicast, (ii) 20\% of the 500 files should be placed at more than 10 users, and 50\% of files should be placed at more than 5 users. Besides, not all the 500 files need to be changed in each update.} The cached files at each user are updated in off-peak time.
This setup is used in the sequel unless otherwise specified.


\vspace{-2.5mm}
\subsection{Impact of Key Parameters}

In the sequel, we analyze the impact of user mobility, user preference similarity parameter $\alpha$, collaboration distance $r_c$, cache size $M$, and Zipf parameter $\beta$ on the offloading probability.

We consider a widely used mobility  model, {\em random walk model}, where a user moves from its current location to a new location by randomly choosing a direction and speed to travel \cite{camp2002survey}. To compute the contact probability matrix, we consider the two-hour period where each user moves 100 seconds before changing direction and speed. The users are initially uniformly distributed, and the speed and direction of each user are uniformly chosen from [$0$, $v_{\max}$] m/s and [$0$, $2\pi$], respectively.  By computing the duration that the $k$th and the $m$th user can establish D2D links, $t^{d}_{k,m}$, in the period of $T_p$ = 2 hours, we can obtain the contact probability $a_{k,m} = \frac{t^{d}_{k,m}}{T_p}$. By increasing $v_{\max}$, users may move with higher speed, and when $v_{max} = 0$, all users keep fixed. In this subsection, both user preference and content popularity are perfect.

\begin{figure}[!htb]
	\centering
\includegraphics[width=0.475\textwidth]{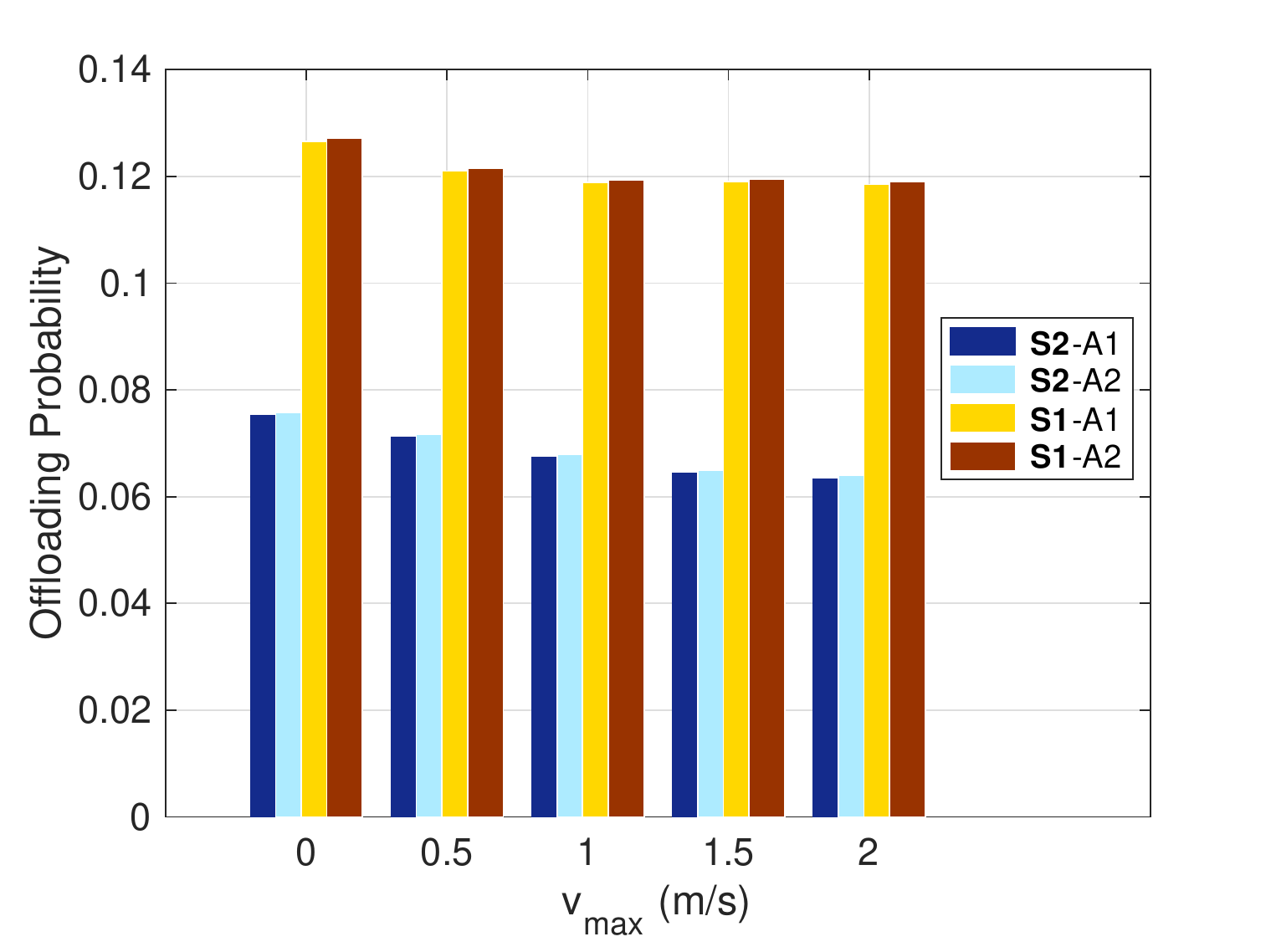}
	\caption{Impact of user mobility, $\alpha =0.36$, $\textbf{S1}$ is the caching policy with user preference, and $\textbf{S2}$ is the policy with content popularity that represents existing caching policy.}\label{fig.3}
\end{figure}

In Fig. \ref{fig.3}, we show the impact of user mobility. ${A1}$ and ${A2}$ in the legend respectively represent the greedy algorithm (i.e., Algorithm 1) and local optimal algorithm (i.e., Algorithm 2), which achieve almost the same offloading probability. The offloading probabilities decrease slightly with the growth of $v_{\max}$, as explained as follows. Owing to the mobility model, the average number of users that a user can establish D2D links with at any time does not change with $v_{\max}$. Then, the total effective cache size seen by the user does not change with $v_{\max}$. On the other hand, every user can contact with more users in the period with higher $v_{\max}$. Then, the caching policy needs to be optimized by considering the preferences of more users, which reduces the cache hit ratio due to heterogeneous user preferences. Since the impact of mobility is not significant, we only consider $v_{max} = 0$ in the sequel.
%
To obtain the results of $A2$ in Fig. \ref{fig.3}, three iterations of step 1 (i.e., $t_{A2} = 3$) is necessary for convergence.
According to analysis in Section \ref{sec:caching}, the time complexity for $A1$ and $A2$ are respectively $O(K^2FM(K^2+1))$ and $O(3KF(K^2+M))$, and $A2$ will be $\frac{KM(K^2+1)}{3(K^2+M)} \approx 167$ times faster than $A1$ when $K=100$ and $M=5$.
Since the proposed local optimal algorithm can achieve the same performance and is faster than the greedy algorithm, we only use $A2$ to obtain the caching policy in the following.

\begin{figure}[!htb]
	\centering
	\subfigure[Offloading probability vs. $\alpha$, $\beta = 0.6$ and $M=5$.]{
		\includegraphics[width=0.475\textwidth]{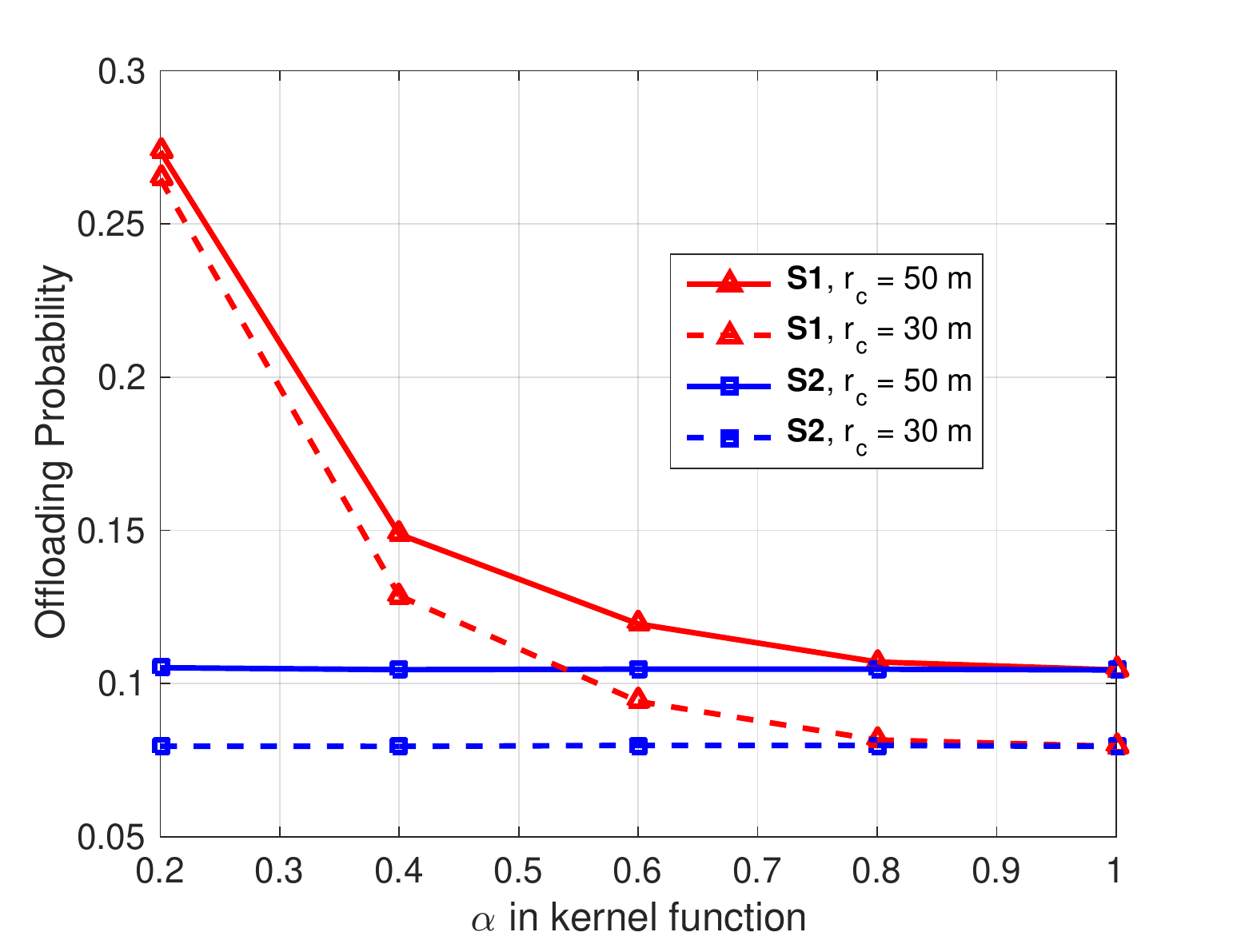}
	}
	\subfigure[Offloading probability vs. $\beta$, $\alpha=0.36$ and $r_c = 30$ m.]{
		\includegraphics[width=0.475\textwidth]{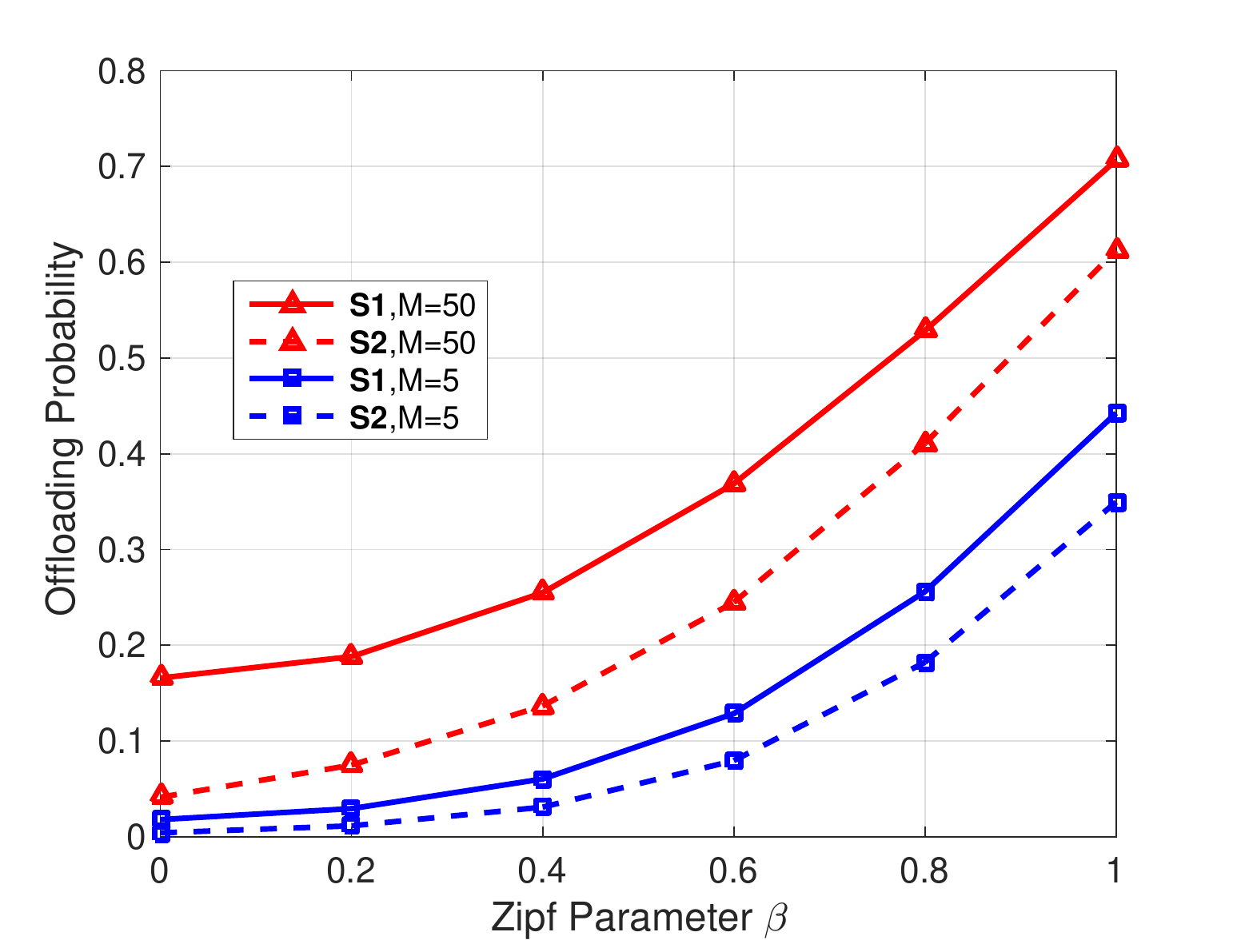}}
	\caption{The impact of $\alpha$, $\beta$, $r_c$ and $M$, $\textbf{S1}$ and $\textbf{S2}$ are caching policies with user preference and content popularity.}\label{fig.4}
\end{figure}

In Fig. \ref{fig.4}(a), we show the impact of $\alpha$ and $r_{\rm c}$. We can see that the offloading gain of \textbf{S1} over \textbf{S2} is high when $\alpha$ is small. This suggests that optimizing caching policy according to user preferences is critical when the user  preferences are less correlated. As expected, when $\alpha \to 1$, the performance of the two policies coincide. The offloading gain is high for large collaboration distance, but the gain by using  \textbf{S1}  reduces as indicated in Remark 3. This is because with the growth of $r_{\rm c}$, the number of users whose preferences a helper should consider for optimizing caching policy increases (when $r_{\rm c} \rightarrow \infty$, the number of users equals to $K$).

In Fig. \ref{fig.4}(b), we show the impact of $\beta$ and $M$. As expected, with the growth of the value of $M$ or $\beta$, the offloading probabilities increase for both $\textbf{S1}$ and $\textbf{S2}$. This is because with a given value of $\alpha$, the preference of every user becomes more skewed when $\beta$ increases.

\subsection{Offloading Gain with Learned User Preference}
In what follows, we demonstrate the performance gain of the proposed caching policy by exploiting user preferences over that with content popularity, and the gain provided by learning with the pLSA model and the priori knowledge. To this end, we compare the following schemes:
\begin{enumerate}
\item ``\textbf{S1-perfect}'': The {\em proposed} caching policy with perfect user preference and active level, which is the solution of problem \textbf{P1}.
\item ``\textbf{S2-perfect}'': The {\em existing} caching policy optimized with perfect content
popularity, which is the solution of problem \textbf{P2} (slightly different from the policies in \cite{JMY.JSAC,chen2017energy}).
\item ``\textbf{S1-EM}'': The {\em proposed} caching policy with $\bf \hat{w}$ and $\bf \hat{Q}$ learned by the EM algorithm.
\item ``\textbf{S2-EM}'': The {\em existing} caching policy with learned local popularity of the $K$ users, which is computed with \eqref{p_f} from the learned user preference by EM algorithm.
\item ``\textbf{S1-prior}'': The {\em proposed} caching policy with $\bf \hat{Q}$ learned by Algorithm \ref{prior_knowledge}.
\item ``\textbf{S2-prior}'': The  {\em existing} caching policy with learned local popularity of the $K$ users, which is computed from the learned user preference by Algorithm \ref{prior_knowledge} as $\hat{p}_f  = \sum_{{\rm u}_k \in \mathcal{U}} \hat P({\rm u}_k , {\rm f}_f )$.
\item ``\textbf{S1-baseline}'': The {\em proposed} caching policy with learned user preference, which is obtained by the ML algorithm without pLSA model.
\item ``\textbf{S2-baseline}'': The  {\em existing} caching policy with learned local popularity of the $K$ users, which is obtained by using the traditional frequency-count popularity prediction method in  \cite{tatar2014survey}. Such a popularity learning method is the same as the method used in \cite{Blasco14learning}.
\end{enumerate}

\begin{figure}[!htb]
	\centering
	\subfigure[$M = 5$.]{
		\includegraphics[width=0.475\textwidth]{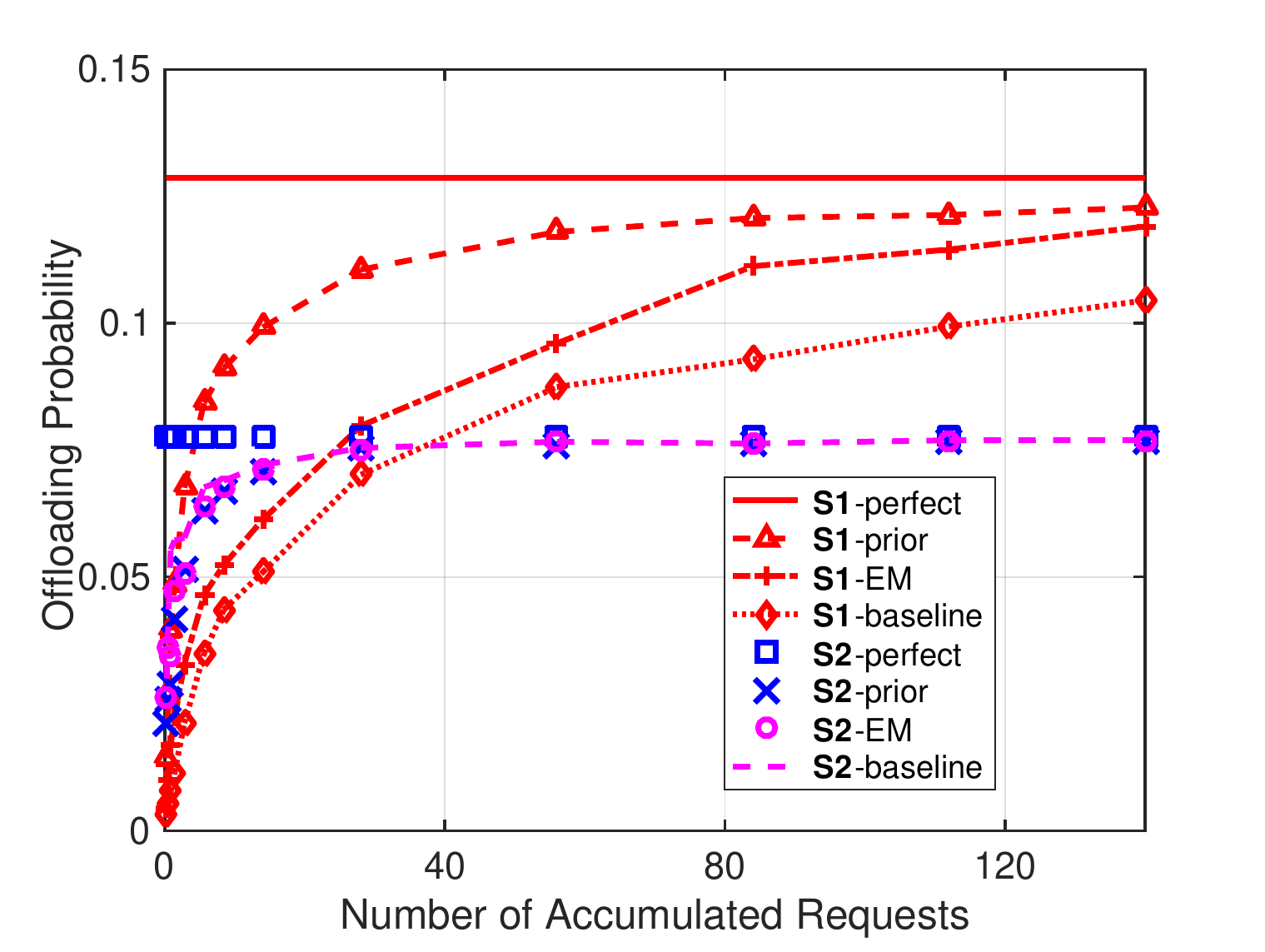}
	}
	\subfigure[$M = 50$.]{
		\includegraphics[width=0.475\textwidth]{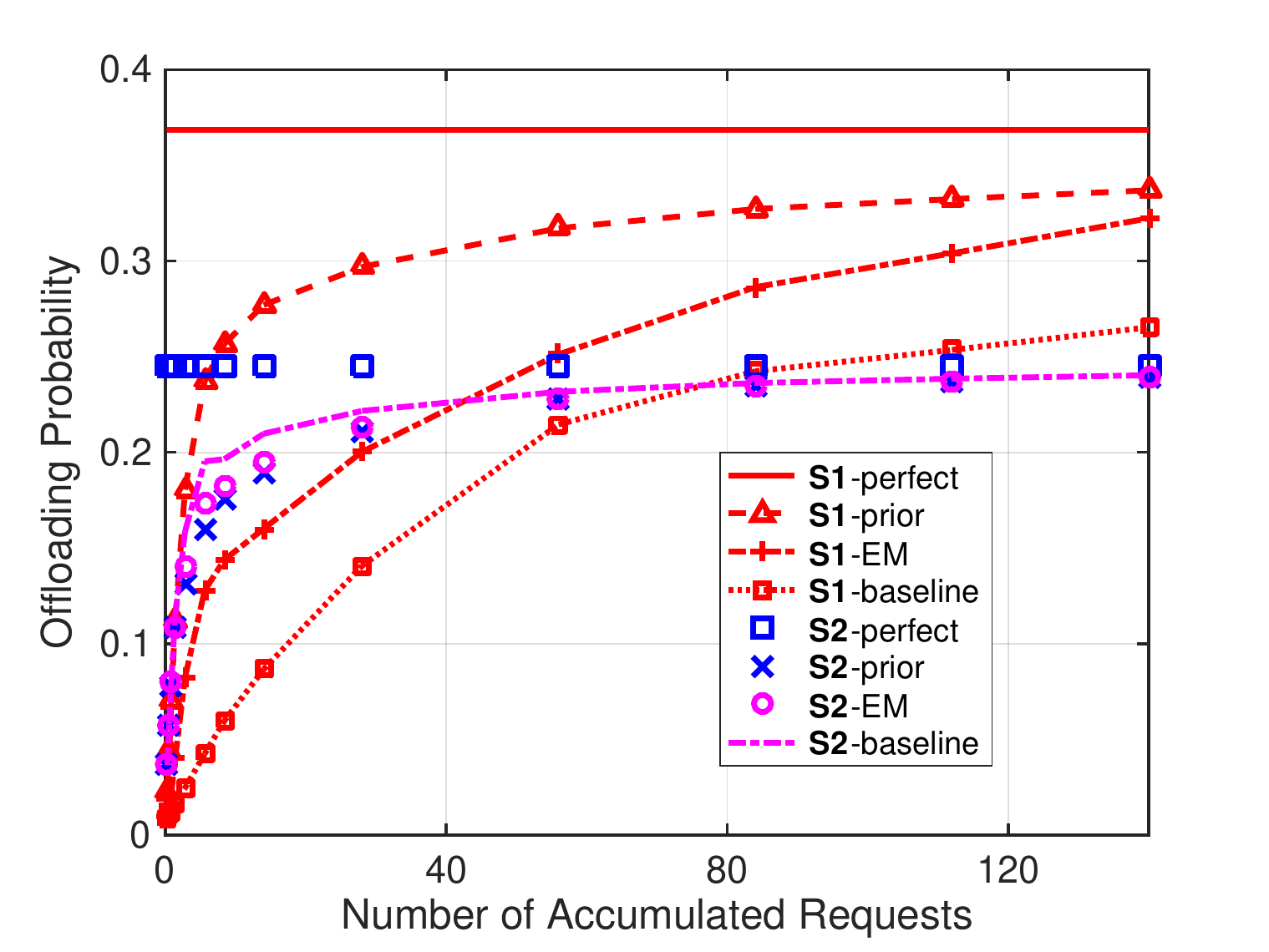}}
	\caption{Convergence performance on synthetic dataset, $\alpha = 0.36$, $r_{\rm c} = 30$ m, $Z=20$ for pLSA. }\label{fig.5}
\end{figure}

In Fig. \ref{fig.5}, we show the offloading probability achieved by these schemes during the learning procedure with the synthetic data. To compare with the results to be obtained with realistic dataset in the sequel, we set the x-axis as the accumulated number of requests. When the request arrival rate of the users in the area is $0.04$ requests per second, which reflects the high traffic load scenario for files with $30$ MBytes size (typical size of the YouTube videos) in \cite{3GPP.PT}, the accumulated 40, 80 and 120 requests correspond to 22.8, 55.6, and 83.3 hours, respectively. It is shown that using pLSA and even the priori information do not help accelerate convergence of local popularity, because the simple frequency-count method already converges rapidly. Compared to the proposed caching policy with learned user preference (\textbf{S1}-EM, \textbf{S1}-prior and \textbf{S1}-baseline), \textbf{S2} with learned local popularity (\textbf{S2}-EM, \textbf{S2}-prior and \textbf{S2}-baseline) converge to \textbf{S2} with perfect  content popularity (\textbf{S2}-perfect) more quickly. This is because the number of requests for each file from each user is much less than that from all the users in the area. Nonetheless, the proposed caching policy with learned user preference can quickly achieve higher offloading probability than \textbf{S2} with learned (and even perfect) content popularity. The proposed caching policy with pLSA (both \textbf{S1}-EM and \textbf{S1}-prior) is superior to the baseline (\textbf{S1}-baseline), especially when the cache size at each user $M$ is large.  This is because some unpopular files will be cached with large $M$. For the unpopular files, the number of accumulated requests is less and user preference learning is more difficult. Besides, we can see that by exploiting prior knowledge of user active level and topic preference, \textbf{S1}-prior converges much faster than  \textbf{S1}-EM.

\begin{figure}[!htb]
	\centering
	\includegraphics[width=0.475\textwidth]{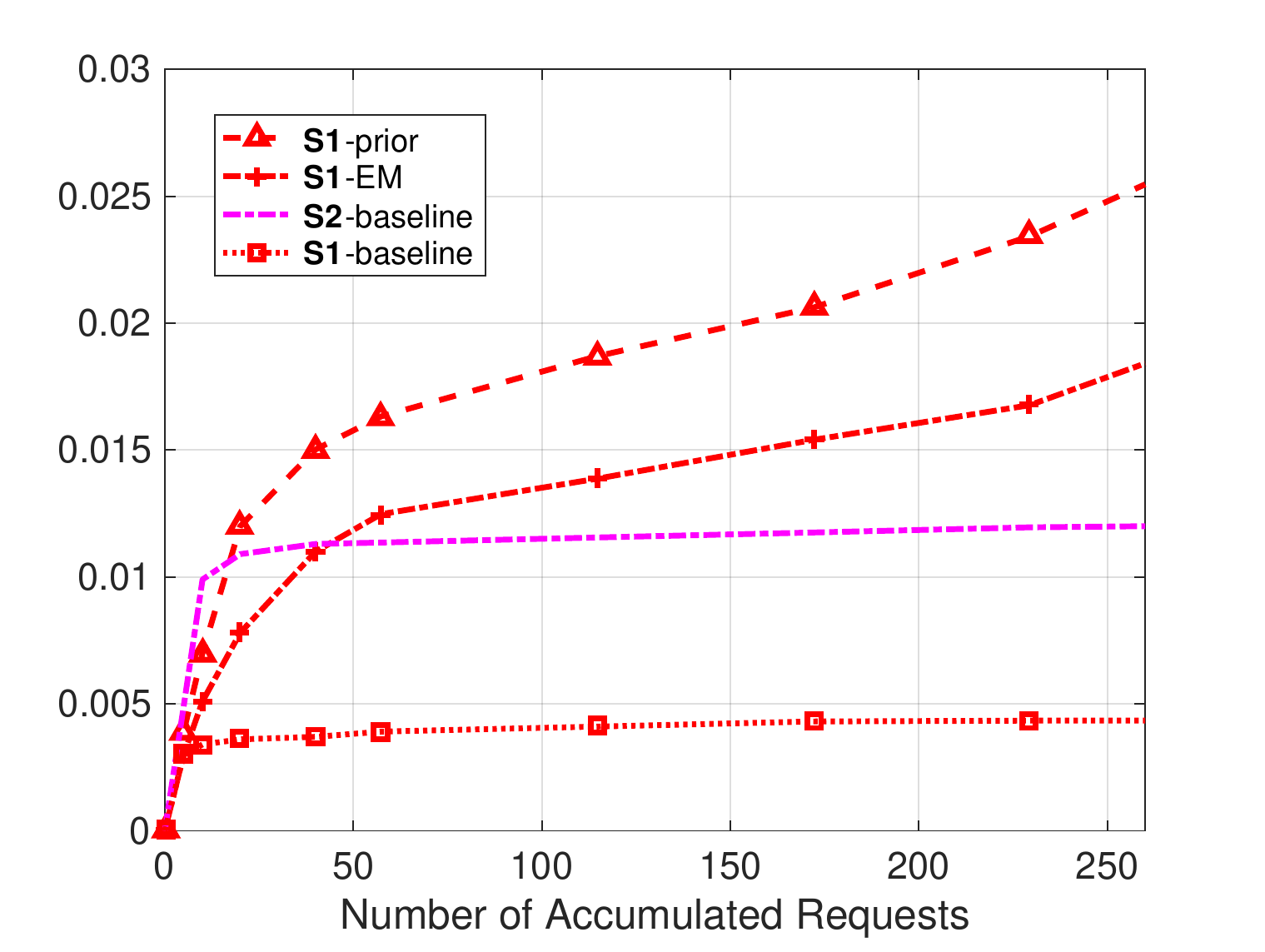}
	\caption{Offloading probability on MovieLens dataset, $r_c=30$ m, $M=5$.}\label{fig.6}
\end{figure}

In Fig. \ref{fig.6}, we show the offloading gain with the MovieLens dataset. Because \textbf{S2}-EM and \textbf{S2}-prior perform closely to \textbf{S2}-baseline, we only simulate \textbf{S2}-baseline here. We randomly choose $100$ users from the dataset (which includes both active and inactive users) and the most popular $3000$ files. The timestamps of user requests are shuffled as in \cite{bacstuug2015big} to ensure the training set has the same user demand statistics as the test set.  Compared with Fig. \ref{fig.5}(a), we can see that the offloading probabilities for all methods on the realistic dataset are less than those with synthetic data. This is because a user requests each movie at most once in the realistic dataset translated from rating data, while a user may request a file more than one time in the synthetic data. In practice, a user may request a file more than once, e.g., the file is a favorite song of the user or an educational video.  Nonetheless, we can see that the proposed caching policy with predicted user preference still achieves much higher offloading gain than existing scheme. Because the prior knowledge of topic preferences is learned on realistic dataset in Fig. \ref{fig.6} rather than perfectly known as in Fig. \ref{fig.5}(a), the performance gain of \textbf{S1}-prior over \textbf{S1}-EM is lower here, which however is still remarkable. Besides, we can see that \textbf{S1}-baseline always performs the worst, because user preference for unvisited files is hard to predict by using the ML algorithm without pLSA model when each user only requests a file at most once. However, we can still predict user preference for the unvisited files with both \textbf{S2}-EM and \textbf{S2}-prior, owing to the pLSA model.

\section{Conclusions}
\label{sec:conclusion}
In this paper, we demonstrated the caching gain by exploiting learned individual user behavior in sending request. We first showed the connection between user preference and content popularity, and provided a probabilistic model to synthesize user preference from content popularity. We then formulated an optimization problem with given user preference and active level to maximize the offloading probability  for cache-enabled D2D communications. Since the problem is NP-hard, a low-complexity algorithm achieving at least 1/2 optimality was proposed to solve the problem. Next, we modeled the user request behavior by pLSA, based on which the EM algorithm was used to learn the user preference and active level. We used a Movielens dataset to analyze several kind of user behavior in requesting contents, and validated the synthetical model. We find that: (i) when the number of users $K$ in an area is large, the file catalog size is a logarithm function of $K$, (ii) the active level of the most active users can be modeled as Zipf distribution, (iii) the preferences for
the most favorable files of each user can be modeled as Zipf distribution, but with different skewness parameters and different file sets, (iv) the user preferences are less similar, and (v) the active level and topic preference of each user change slowly over time, say in the time scale of year. Based on the 5th observation from the real dataset, we introduced a prior knowledge based algorithm to exploit the active level and topic preference previously learned, which shows the potential of transfer learning. Simulation results showed that using pLSA can quickly learn the individual user behavior, and the prior knowledge based algorithm converges even faster. Compared to existing caching policy using content popularity, the performance can be remarkably improved by the caching policy exploiting user preferences, both on the synthetic data with parameters fitted from real dataset and on the MovieLens dataset.

\appendices
\numberwithin{equation}{section}

\section{Proof of Proposition \ref{proposition3}}
\label{a:3}

The objective function of problem $\textbf{P1}'$ can be further derived as
\begin{equation}
\label{equ.obj.opt6}
\begin{aligned}\textstyle
f_{\rm off}({\bf c}_{k'} ) & \textstyle =   \sum_{k=1}^{K} w_k \sum_{f=1}^{F} q_{f|k} \left(1-\prod_{m=1,m\neq k'}^{K} (1- a_{k,m}c_{m,f}) (1- a_{k,k'}c_{k',f}) \right) \\
& \textstyle =   \underbrace{1- \sum_{f=1}^{F} \sum_{k=1}^{K} w_k  q_{f|k} \prod_{m=1,m\neq k'}^{K} (1- a_{k,m}c_{m,f})}_{(a)}  \\
& \textstyle  + \sum_{f=1}^{F} c_{k',f} \textstyle \underbrace{\left(\sum_{k=1}^{K} w_k  q_{f|k} a_{k,k'}\prod_{m=1,m\neq k'}^{K} (1- a_{k,m}c_{m,f})\right)}_{(b)},
\end{aligned}
\end{equation}
where both terms in (a) and (b) are not related to $c_{k',f}$. Then, solving the problem in \eqref{equ.opt6} is equivalent to solving the following problem
\begin{equation}
\label{equ.opt7}
\begin{aligned} \textstyle
\textbf{P1}' \quad&    \textstyle   \max_{c_{k',f}}\,\, &&   \textstyle  \sum_{f=1}^{F} c_{k',f} {\left(\sum_{k=1}^{K} w_k q_{f|k} a_{k,k'} \prod_{m=1,m\neq k'}^{K} (1- a_{k,m}c_{m,f})\right)} \stackrel{(a)}{=}  \sum_{f=1}^{F} c_{k',f} b_{k',f}  \\
&   \textstyle  s.t.  &&   \textstyle \sum_{f=1}^{F} c_{k',f} \leq M, c_{k',f} \in \{0,1\}, 1 \leq f \leq F,
\end{aligned}
\end{equation}
where (a) is obtained by letting $b_{k',f} = \sum_{k=1}^{K} w_k  q_{f|k} a_{k,k'}\prod_{m=1,m\neq k'}^{K} (1- a_{k,m}c_{m,f})$. By finding file indices of the maximal $M$ values of $b_{k',f} (1\leq f \leq F)$ to constitute the set $\mathcal{I}_{k'}$, it is not hard to show that the optimal caching policy $c^*_{k',f}$ can be obtained as \eqref{equ.res.algo}.

To obtain $c^*_{k',f}$, we need to compute $b_{k',f}$ with time complexity $O(K^2F)$ and then choose the  maximal $M$ values of $b_{k',f}$ with complexity $O(FM)$. Finally, we can prove that the optimal solution of problem \eqref{equ.opt6} can be obtained with complexity $O(K^2F+FM) = O(F(K^2+M))$.

\section{Proof of Proposition \ref{proposition4}}
\label{a:4}

Denote the caching policy at the $k'$th user after the $(t-1)$th iteration as ${\bf c}^{(t-1)}_{k'}$. In the $t$th iteration, the offloading probability before step \ref{opt_c_k} of Algorithm \ref{local_opt_algo} is $f_{\rm off}({\bf c}^{(t-1)}_{k'} )$ as in \eqref{equ.obj.opt6}. After that step, ${\bf c}^{(t)}_{k'}$ is computed for the $k'$th user and the corresponding offloading probability is $f_{\rm off}({\bf c}^{(t)}_{k'} )$. By subtracting $f_{\rm off}({\bf c}^{(t-1)}_{k'} )$ from $f_{\rm off}({\bf c}^{(t)}_{k'} )$, we can obtain

\begin{equation}
\label{equ.subtrac}
\begin{aligned}\textstyle
 \textstyle f_{\rm off}({\bf c}^{(t)}_{k'} ) - f_{\rm off}({\bf c}^{(t-1)}_{k'} ) & \textstyle =   \sum_{k=1}^{K} w_k \sum_{f=1}^{F} q_{f|k} \left(1-\prod_{m=1,m\neq k'}^{K} (1- a_{k,m}c_{m,f}) (1- a_{k,k'}c^{(t)}_{k',f}) \right) \\ &\textstyle- \sum_{k=1}^{K} w_k \sum_{f=1}^{F} q_{f|k} \left(1-\prod_{m=1,m\neq k'}^{K} (1- a_{k,m}c_{m,f}) (1- a_{k,k'}c^{(t-1)}_{k',f}) \right)  \\
  & \textstyle\stackrel{(a)}{=}     \sum_{f=1}^{F} c^{(t)}_{k',f} b_{k',f} - \sum_{f=1}^{F} c^{(t-1)}_{k',f} b_{k',f} \\
& \textstyle \stackrel{(b)}{=}  (\max_{c_{k',f}}  \sum_{f=1}^{F} c_{k',f} b_{k',f} )- \sum_{f=1}^{F} c^{(t-1)}_{k',f} b_{k',f} \geq 0,
\end{aligned}
\end{equation}
where (a) is obtained by substituting \eqref{b_k_f} and \eqref{equ.obj.opt6}, and (b) is obtained from \eqref{equ.opt7}. Thus, the offloading gain is monotonically improved until convergence.

To prove the optimality guarantee of the local optimal algorithm, we first convert the offloading probability into a function of a set instead of a matrix (i.e., ${\bf C}$). Denoting ${\rm f}_{f}^{k}$ as an action that caching the $f$th file at the $k$th user. Recall that $c_{k,f}=1$ represents the $k$th user caching the $f$th file. Then, the caching policy for the $k$th user, ${\bf c}_k = [c_{k,1},c_{k,2},...,c_{k,F}]$, can be re-expressed as a set $\mathcal{C}_k = \{{\rm f}^{k}_f|c_{k,f}=1\}$, i.e., caching which files at the $k$th user. Let $\mathcal{C} = \{\mathcal{C}_1, \mathcal{C}_2, ..., \mathcal{C}_K \}$, then problem \textbf{P1} is equivalent to the following problem,
\begin{equation}
\label{equ.opt4}
\begin{aligned}\textstyle
\quad&   \textstyle  \max_{\mathcal{C}}\,\, &&  \textstyle f_{\rm }(\mathcal{C}) =  \sum_{k=1}^{K} w_k \sum_{f=1}^{F} q_{f|k}\left(1-\prod_{{\rm f}_{f}^{m} \in \mathcal{C}} (1- a_{k,m}) \right) \\
& \textstyle  s.t.  && \textstyle   |\mathcal{C}_k| \leq M, 1 \leq k \leq K.
\end{aligned}
\end{equation}

By defining a set $\mathcal{S} = \{  {\rm f}_{1}^{1}, {\rm f}_{2}^{1}, ... , {\rm f}_{F}^{1}, ..., {\rm f}_{1}^{K},{\rm f}_{2}^{K},...,{\rm f}_{F}^{K} \}$, we can see that $\mathcal{C} \subseteq \mathcal{S}$ and $f_{\rm}(\mathcal{C}): 2^{\mathcal{S}} \rightarrow R$ is a discrete set function on subsets of $\mathcal{S}$.
Let $\mathcal{A},\mathcal{B} \subseteq \mathcal{S}$, $\mathcal{A} \subseteq \mathcal{B}$, and ${\rm f}_{k'}^{f'} \in \mathcal{S} \setminus \mathcal{B}$.

Denote the global optimal caching policy as $\mathcal{C}^* = \{\mathcal{C}^*_1,\mathcal{C}^*_2,...,\mathcal{C}^*_K\}$, a local optimal caching policy obtained by Algorithm \ref{local_opt_algo} as $\mathcal{C}^L = \{\mathcal{C}^L_1,\mathcal{C}^L_2,...,\mathcal{C}^L_K\}$, and caching policy at users except the $k$th user as $\bar{\mathcal{C}^L_k} = \{\mathcal{C}^L_1,\mathcal{C}^L_2,.,\mathcal{C}^L_{k-1},\mathcal{C}^L_{k+1},..,\mathcal{C}^L_K\}$.
Then, we can obtain
\begin{equation}
\begin{aligned}\textstyle
  & f(\mathcal{C^*}) - f(\mathcal{C}^L) \textstyle \stackrel{(a)} {\leq} f(\mathcal{C^*} \cup\mathcal{C}^L ) - f(\mathcal{C}^L) \triangleq  f_{\mathcal{C}^L}(\mathcal{C^*}) \stackrel{(b)}{\leq} \sum_{k=1}^{K}f_{\mathcal{C}^L}(\mathcal{C}^*_k) \\
 & \textstyle \stackrel{(c)}{\leq}  \sum_{k=1}^{K}f_{\bar{\mathcal{C}^L_k}}(\mathcal{C}^*_k) \stackrel{(d)}{\leq}  \sum_{k=1}^{K}f_{\bar{\mathcal{C}^L_k}}(\mathcal{C}^L_k)  \textstyle \stackrel{(e)}{\leq}  f(\mathcal{C}^L_1)  +\sum_{k=2}^{K}f_{\cup_{i=1}^{k-1}\mathcal{C}^L_i}(\mathcal{C}^L_k) = f(\mathcal{C}^L),
\end{aligned}
\end{equation}
where (a) is obtained because offloading probability is a monotone increasing function, i.e., $f(\mathcal{C}^*) \leq f(\mathcal{C^*} \cup \mathcal{C}^L)$, (b) is obtained by the property that for set $\mathcal{A},\mathcal{B},\mathcal{C} \subseteq \mathcal{S}$, we have $f_\mathcal{A}(\mathcal{B} \cup \mathcal{C}) \leq f_\mathcal{A}(\mathcal{B}) + f_\mathcal{A}(\mathcal{C})$, (c) and (e) are obtained by the property that for set $\mathcal{C} \subseteq \mathcal{A} \subseteq \mathcal{S}$ and $\mathcal{B} \subseteq \mathcal{S}$, $f_\mathcal{A}(\mathcal{B} ) \leq f_\mathcal{C}(\mathcal{B})$, and (d) is obtained because for any caching policy at the $k$th user denoted as $\mathcal{C}^a_k$, we have $f_{\bar{\mathcal{C}^L_k}}(\mathcal{C}^L_k) - f_{\bar{\mathcal{C}^L_k}}(\mathcal{C}^a_k) = f(\mathcal{C}^L_k \cup {\bar{\mathcal{C}^L_k}}) - f(\mathcal{C}^a_k \cup {\bar{\mathcal{C}^L_k}}) \geq 0$ considering $\mathcal{C}^L$ is the local optimum of Algorithm \ref{local_opt_algo}. Thus, we have $f(\mathcal{C}^L) \geq \frac{1}{2} f(\mathcal{C^*})$, and Proposition \ref{proposition4} follows.
\bibliographystyle{IEEEtran}
\bibliography{CBQ_J}

\end{document}